\documentclass[
12pt,
amsmath,
superscriptaddress,
floatfix,
notitlepage,
longbibliography,
]{revtex4-1}

\usepackage{graphicx}
\usepackage{amsmath}
\usepackage{comment}

\begin{document}

\title{The Capacitance-Power-Hysteresis Trilemma in Nanoporous Supercapacitors}

\author{Alpha A Lee }
\affiliation{School of Engineering and Applied Sciences, Harvard University, Cambridge, MA 02138, USA} 
\affiliation{Mathematical Institute, Andrew Wiles Building, University of Oxford, Woodstock Road, Oxford OX2 6GG, United Kingdom}

\author{Dominic Vella}
\affiliation{Mathematical Institute, Andrew Wiles Building, University of Oxford, Woodstock Road, Oxford OX2 6GG, United Kingdom}

\author{Alain Goriely}
\affiliation{Mathematical Institute, Andrew Wiles Building, University of Oxford, Woodstock Road, Oxford OX2 6GG, United Kingdom}

\author{Svyatoslav Kondrat}
\affiliation{IBG-1: Biotechnology, Forschungszentrum J\"{u}lich, 52425 J\"{u}lich, Germany}
\affiliation{Department of Chemistry, Faculty of Natural Sciences, Imperial College London, SW7 2AZ, UK}

\begin{abstract}
Nanoporous supercapacitors are an important player in the field of energy storage that fill the gap between dielectric capacitors and batteries. The key challenge in the development of supercapacitors is the perceived tradeoff between capacitance and power delivery. Current efforts to boost the capacitance of nanoporous supercapacitors focus on reducing the pore size so that they can only accommodate a single layer of ions. However, this tight packing compromises the charging dynamics and hence power density. We show via an analytical theory and Monte Carlo simulations that charging is sensitively dependent on the affinity of ions to the pores, and that high capacitances can be obtained for ionophobic pores of widths significantly larger than the ion diameter.  Our theory also predicts that charging can be hysteretic with a significant energy loss per cycle for intermediate ionophilicities. We use these observations to explore the parameter regimes in which a capacitance-power-hysteresis \emph{trilemma} may be avoided.
\end{abstract}

\makeatother
\maketitle

\section{Introduction}


The physics of charge storage at the nanoscale has received significant attention in recent years due to its relevance for efficient energy storage and the development of novel green technologies~\cite{lu2013supercapacitors, simon2012capacitive, simon2008materials}. In particular, extensive effort has been channeled into studying electrical double layer capacitors (also called supercapacitors) in which energy is stored at the electrode-electrolyte interface. Their importance for energy storage has stimulated the development of novel techniques for fabrication of conducting nanoporous materials~\cite{frackowiak:07, simon2008materials}. For instance, high-temperature chlorination of carbides produces carbon electrodes with a network of slit and/or cylindrical nanopores with narrow pore-size distribution about a well-controlled average pore size ~\cite{gogotsi:natmat:03}. There is also an emerging class of graphene-based electrodes consisting of aligned slit nanopores with pore sizes comparable to the ion diameter~\cite{yoo:nanolett:graphene:11, yang:sci13:graphene}. Additionally it is possible to manipulate the ion-pore interactions by functionalising carbons. For instance, preparing carbon nanofibers in the presence of potassium hydroxide changes the surface functionality and increases the ion-pore attraction~\cite{richey:jpcc:philicity:14}.

\begin{figure}
\centering
\includegraphics[scale=0.3]{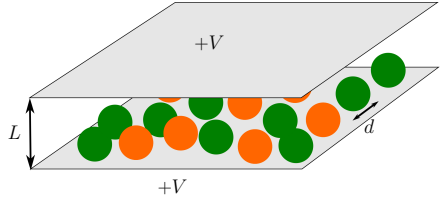}
\caption{Schematic drawing of the model porous electrode under consideration: Ions of diameter $d$ are confined between two metallic surfaces separated by a distance $L$. A potential $V$ (relative to the bath) is applied.}
\label{fig:model}
\end{figure}

Nanoporous supercapacitors benefit from high surface-to-volume ratio of these materials with an increase in the volumetric capacitance observed as the surface area of the electrode increases~\cite{conway1999electrochemical}. However, pioneering experiments \cite{pinero:carbon:06, chmiola2006anomalous,largeot2008relation,lin2009solvent} have shown that a drastic increase in \emph{surface}-specific capacitance is achieved when the average pore size approaches the ion diameter. Using a model of a single \emph{metallic} slit-shaped nanopore (\textit{c.f.}~Figure~\ref{fig:model}), this `anomalous' increase of capacitance has been explained by the emergence of a `superionic state'  in which the inter-ionic interactions become exponentially screened. Decreasing the pore size promotes screening which decreases the energy penalty for packing like charges and unbinding ion pairs, purportedly leading to an increase in the capacitance~\cite{kondrat2011superionic}. This reasoning applies also to non-perfect metals~\cite{Rochester2013, schmickler:ec:14, schmickler:ea:15, vella2016quantum}. In particular, recent quantum density-functional calculations have shown that the ion-ion interactions are exponentially screened in carbon nanotubes~\cite{schmickler:ec:14, schmickler:ea:15} and it is reasonable to expect a similar behaviour for other types of confinement, including slit pores. 




This effect of metallic screening has been observed in molecular dynamics simulations that use more elaborate models to account for complex pore geometries~\cite{merlet2012molecular,merlet2013highly} and realistic ions~\cite{wu2011complex, qiao2012voltage, vatamanu2013increasing}. If metallic screening is the sole driver of increased capacitance, the capacitance can only be optimized when the pore size equals the ion size. However, such a close-fitting pore is detrimental to the charging dynamics \cite{mysyk:ec:09} because of the reduced effective diffusivity~\cite{kondrat:jpcc:13, lee:nanotech:14, kondrat:nm:14} and because the kinetic barrier to pore entry is large. Increased capacitance therefore appears to come at the cost of prolonged charging. This leads to a dilemma in the design of supercapacitors --- should the design be chosen to optimize capacitance or power? Alternatively, one might naturally ask: Can the capacitance be maximized away from strerically close-fitting pores? 

It is known that charging of flat electrodes may show a \emph{hysteresis} \cite{lockett:jpcc:hyster:08, zhou:ec:hysterflat:10, uysala:jpcc:hyster:13,rotenberg2015structural}, \emph{i.e.}~the differential capacitance depends on the initial voltage and the direction of scan~\cite{lockett:jpcc:hyster:08, druechler:jpcc:hyster:10}. This is connected with the existence of two or more metastable states of an ionic liquid at the electrode surface \cite{rotenberg2015structural}. In nanoconfinement, there is evidence from a mean-field study~\cite{kondrat2011superionic}, Monte Carlo~\cite{kiyohara2011phase} and molecular dynamics~\cite{xing:jpcl:trans:13,vatamanu:acsnano:15} simulations that charging of slit nanopores can proceed via a voltage--induced discontinuous transition that is manifested by an abrupt change in the ion density. Such discontinuous transitions can be detrimental to the operation of supercapacitors because of \emph{hysteretic energy losses} when the charging and discharging routes follow different metastable branches. It is thus important to know whether (and when) charging is hysteretic, and how hysteresis might be avoided altogether. This adds another dimension to the dilemma already mentioned.

To answer these questions, we combine a mean field theory with Monte Carlo simulations for a model slit nanopore (Figure~\ref{fig:model}). We consider monovalent ions and a single slit-shaped metallic nanopore. The pore entrance and closing are ignored, and charging is modelled by applying a non-zero potential to the pore walls. This or similar models have previously been used to study charge storage~\cite{kondrat2011superionic, kondrat2011superionicMC, kiyohara2011phase, kiyohara2012phase, kiyohara2013phase, jiang2011oscillation, vatamanu2013increasing, jiang:nanoscale:14} and the dynamics of charging~\cite{kondrat:jpcc:13, kondrat:nm:14} of nanoporous supercapacitors. Here, we focus specifically on pores whose sizes are comparable with the ion diameter. In this limit, the system is quasi two-dimensional, and we can assume that ions are located on the central symmetry plane of the pore. This allows us to develop a mean-field theory in two dimensions, whereby, improving on a model developed previously \cite{kondrat2011superionic}, the entropy of out-of-plane packing of ions (for $L > d$) is taken into account by introducing an effective pore-width dependent ion diameter $d^* \le d$ (see Appendix~\ref{app:mft}; we neglect the out-of-plane effects for the electrostatic interactions as they are subdominant). We supplement and compare our mean-field results with grand canonical Monte Carlo simulations of the same system in three dimensions (for simulation details see Appendix~\ref{app:mc}).

The model employed here ignores the chemical details of ions and the complex structures of nanoporous electrodes. With these deliberate simplifications we aspire to reveal and understand the essential physics at work, without the complexity of real supercapacitors. In particular, our study reveals that contrary to the long-standing paradigm, a maximal capacitance can actually be achieved when pores are appreciably \emph{wider} than the ion diameter. In addition, we show that there is, in general, charging hysteresis with significant energy loss per cycle. However, this hysteresis can be evaded by carefully tuning the ion-pore interaction energy. This study therefore reveals generic features of such systems that we believe should apply more generally and hence may provide a framework within which to design optimal nanoporous supercapacitors and avoid the capacitance-power-hysteresis trilemma discussed above.



\section{Ionophobicity of pores}
\label{sec:ionophobicity}

Our primary interest is to determine how the affinity of ions towards pores affects charging. In our model, this affinity is controlled by the electrochemical potential, 
\begin{equation}
h_\pm = \pm eV + \delta E_\mathrm{self} + \delta E_\pm,
\end{equation} 
where $V$ is the applied voltage, $e$ the elementary charge, $\delta E_\mathrm{self}$ the ions' self-energy (see Eq.~(\ref{eq:self}) in Appendix~\ref{app:mft:pot}), and $\delta E_\pm$ the resolvation energy \cite{kondrat2011superionic}. The resolvation energy is the energy of transferring an ion from the bulk to the pore, and includes here the chemical potential of ions in the bulk. We shall assume $\delta E_+ = \delta E_- = \delta E$ for simplicity.
 
\begin{figure}
\centering
\includegraphics[width=.45\textwidth]{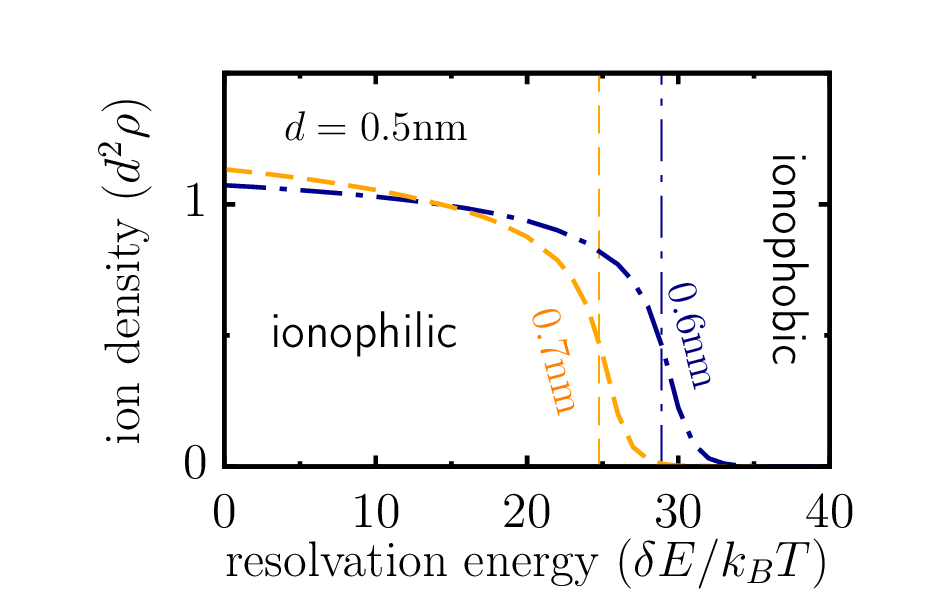} 
\caption{An ion's resolvation energy $\delta E$ determines the ion density (per surface area) inside nanopores at no applied voltage. At large $\delta E$ ions prefer to stay outside the pores, the pores are (nearly) free of ions; we term such pores \emph{ionophobic}. In the opposite case of small or negative $\delta E$ the pores are occupied by ions and we call them \emph{ionophilic}. The crossover between ionophilic and ionophobic occurs when $\delta E_\textrm{crossover} = l_B \ln(2)/L$, where $l_B$ is the Bjerrum length and $L$ pore width (see text); $\delta E_\textrm{crossover}$ for two pore sizes ($L=0.6$nm and $L=0.7$nm) are shown by thin vertical lines. Pores with $\delta E$ close to $\delta E_\textrm{crossover}$ have moderate ion densities at zero applied voltages and we term them \emph{weakly ionophilic/ionophobic}. The ion diameter is $d=0.5$nm and the Bjerrum length $l_B=25$nm. The plots have been obtained by using the mean-field theory (see Appendix~\ref{app:mft}).
}
\label{fig:phobicity}
\end{figure} 

At zero applied voltage, $h_\pm = h_0 = \delta E_{\mathrm{self}} + \delta E$ and the sign of $h_0$ determines whether the ion-pore interactions are favourable. For pores with $h_0>0$ (unfavourable ion-pore interaction), the occupancy of the pore at zero applied voltage is expected to be low: we therefore refer to such pores as \emph{ionophobic}. For large positive $h_0$ the pore will be (almost) completely free of ions at zero applied voltage: we therefore refer to such pores as \emph{strongly ionophobic}. Conversely pores with $h_0<0$ will be termed \emph{ionophilic}, and a large negative $h_0$ will correspond to \emph{strongly ionophilic} pores, which are nearly fully occupied by ions at no applied voltage. $h_0=0$ marks the crossover between \emph{ionophilic} and \emph{ionophobic} behaviours, which therefore occurs at a resolvation energy $\delta E = \delta E_{\mathrm{crossover}} = - \delta E_\mathrm{self} = l_B \ln(2)/L$ (in units of $k_BT$) where $l_B$ is the Bjerrum length (see Appendix~\ref{app:mft:pot} for an estimate of $\delta E_{\mathrm{self}}$). For instance, for  a typical Bjerrum length of $25$nm and a $0.7$nm wide pore we obtain $ \delta E_{\mathrm{crossover}} \approx  25 \; k_BT$, while for a $0.6$nm wide pore we have $ \delta E_{\mathrm{crossover}} \approx  29\; k_BT$ (vertical lines in Figure~\ref{fig:phobicity}).

To obtain an estimate for the resolvation energy, we decompose $\delta E$ into $\delta E \approx \delta E_{\mathrm{desolv}} + \delta E_{\mathrm{non-ele}}$, where $\delta E_{\mathrm{desolv}}$ is the desolvation energy of the bulk ionic liquid (transferring one solvated ion from the bulk liquid to the vapour state), and $\delta E_{\mathrm{non-ele}}$ is the ion-pore non-electrostatic interactions (ion-pore electrostatic interactions are accounted for in $\delta E_\mathrm{self}$). A combination of quantum mechanical density functional calculations and molecular dynamics simulations~\cite{jover2014screening} suggests $\delta E_{\mathrm{desolv}} \approx 65-110 k_B T $ per ion~\footnote{From ref.~\cite{jover2014screening}, the energy to transfer an ion pair from the bulk to vacuum $\approx 50-85 k_B T$, and the energy to dissociate an ion pair at vacuum $\approx  80-140 k_B T$, therefore the desolvation energy of a single ion $\approx 65 -115k_BT$.}. The main source of non-electrostatic interactions is the van der Waals attraction, of magnitude $\delta E_{\mathrm{vdW}} \approx - 70 k_B T$~\footnote{Estimate obtained from Ref.~\cite{jover2014screening}, where a cylindrical geometry is studied. Our estimate is obtained by taking the large radius asymptotic value of the ion-wall van der Waals interaction energy, \emph{i.e.} $E_{\mathrm{vdw,slit}} \approx 2  E_{\mathrm{vdw,cylinder}} (R \rightarrow \infty)$, as the $R \rightarrow \infty$ asymptotic value of the van der Waals interaction energy corresponds to the interaction energy between an ion and one side of the pore.}. We thus find that, roughly, $ -5 \lesssim \delta E/(k_B T) \lesssim 45$ (note that the crossover between ionophobic and ionophilic pores, $\delta E_{\mathrm{crossover}}$, lies in this range).  We stress that this range is not exhaustive as other physical effects (such as specific surface chemistry of the pore or the presence of solvent) have not been taken into account in this analysis. In particular, recent experimental~\cite{griffin:natmat:15, force:jacs:NMRDyn} and theoretical~\cite{rochester:1d} studies suggest that ionophobicity can be effectively controlled by changing the solvent concentration. In any case, we emphasize that the definitive metric of ionophilicity/-phobicity is the occupancy of the pore at zero applied voltage, as shown in Figure~\ref{fig:phobicity}.


\section{Searching for maximal capacitance}

The key quantity characterizing the low voltage capacitance response of a supercapacitor is the differential capacitance at zero voltage,
\begin{equation}
C_{D}(0) = \frac{\mathrm{d} Q}{\mathrm{d}V} \bigg|_{V=0},
\label{CD_at_0V}
\end{equation} 
where $Q$ is the charged stored in a pore, and $V$ is the applied potential. In the following discussion, we compute $Q(V)$ by minimising a mean-field free energy function or directly from Monte Carlo simulations (as discussed in Appendix~\ref{app:mc}). Clearly, $Q(0)=0$ by electroneutrality, while $C_{D}(0)$ describes the response of the system to an applied voltage and is in general nonzero.


The mean-field approximation for the Helmholtz free energy is given by 
\begin{equation}
\label{eq:fe}
\beta F = U_{\rm el}(\rho_+,\rho_-)  - S(\rho_+,\rho_-) + \sum_{\alpha = \pm} h_\alpha \rho_\alpha, 
\end{equation}
where $\beta=(k_BT)^{-1}$ (with $k_B$ being the Boltzmann constant and $T$ temperature) and $\rho_\pm$ is the two-dimensional density of $\pm$ ions. $k_BTU_{\rm el}$ is the contribution to the free energy due to electrostatic interactions, $k_BTS$ is the excluded volume entropic contribution, and we assume that the density of ions in the slit pore is homogeneous (see Appendix~\ref{app:mft} for expressions of $U_{\rm el}$ and $S$). To obtain $Q(V)$ and hence $C_{D}(0)$, we minimize $F$ over $\rho_\pm$ subject to fixed $V$, noting that the charge per unit area is $Q = e(\rho_+-\rho_-)$, where $e$ is the elementary charge. 

It is possible that the free energy has two minima, one of which can be metastable, and charging and discharging may follow different metastable branches. This possibility will be discussed in Section ~\ref{hyster_sect}. We will first focus on the parameter regime where the free energy minimum is unique.

\begin{figure*}
\centering
\includegraphics[width=0.9\textwidth]{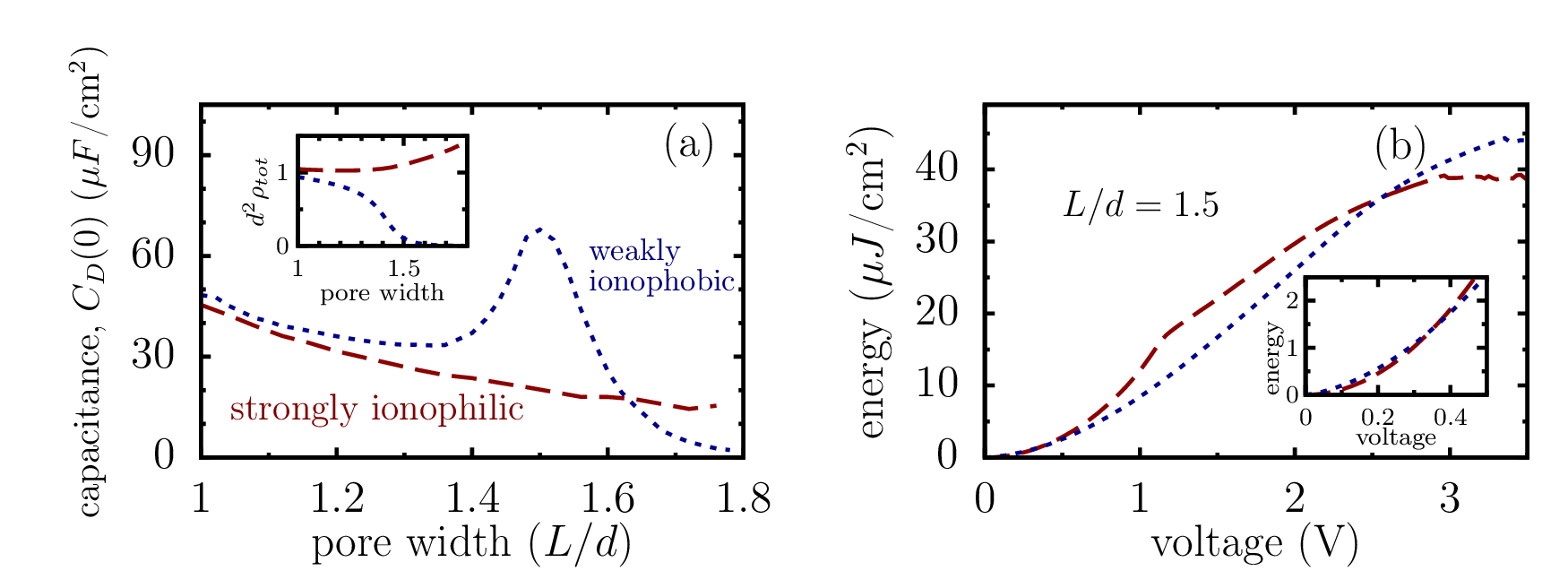}
\caption{
(a)~Differential capacitance at zero voltage $C_D(0)$ as a function of pore width and (b) the stored energy as a function of voltage (for $L=0.75$nm) from the mean field theory, plotted for weakly ionophobic ($\delta E= 25k_BT$) and strongly ionophilic ($\delta E= 10k_BT$) pores. The ion diameter is $0.5$nm. The inset in (a) shows how the total ion density at zero voltage depends on pore width. The inset in (b) highlights the stored energy at low voltages.
}
\label{fig:capen_pore}
\end{figure*}

For strongly ionophilic pores, we find that the capacitance is maximal for the smallest pores possible, in agreement with the conventional view that the capacitance increases monotonically as the pore size decreases~\cite{pinero:carbon:06, chmiola2006anomalous,largeot2008relation,lin2009solvent} (dashed line in Figure~\ref{fig:capen_pore}(a)). Surprisingly, however, we find that for weakly ionophobic pores the differential capacitance has a global maximum when the pore width is significantly larger than the ion diameter (though still smaller than $2d$). This behaviour is due to two competing effects: On the one hand, the loss of ion-image interactions and an increase in electrostatic interactions hinders ions from entering the pore. On the other hand, the same factors also render the pore less populated at zero voltage (the inset in Figure \ref{fig:capen_pore}a). As the width varies, a peak is achieved when the decrease in the total density frees up enough space in the pore that counter-ion insertion becomes entropically favourable. In fact, for narrow and weakly ionophilic pores the charging at low voltages is dominated by swapping coions for counterions and expelling coions; for wider pores, it is the counterion insertion that drives charging.

This charging behaviour is in contrast to strongly ionophilic pores, where total ion density increases with increasing pore width because the out-of-plane degrees of freedom allow ions to pack more efficiently. Therefore, both entropy and ion-ion as well as ion-image interactions work against charging as the pore size ($L$) increases, and so the capacitance decreases monotonically for increasing $L$. The case of strongly ionophobic pores is not considered here --- charging commences only when ions can overcome the ionophobicity barrier ($eV \approx \delta E$) and hence the capacitance at zero voltage is low or vanishing.

\begin{figure*}
\centering
\includegraphics[width=0.9\textwidth]{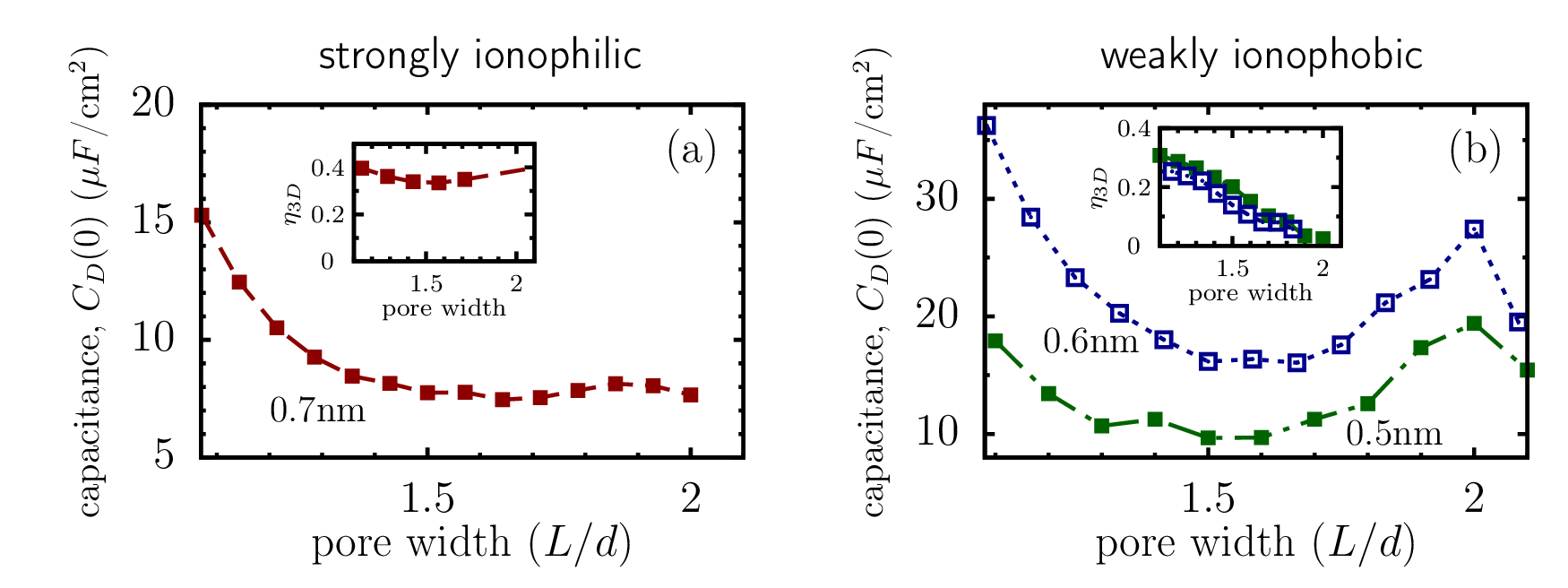}
\caption{The calculated dependence of the differential capacitance, $C_D(0)$, as the pore width varies. The results of Monte Carlo simulations are shown for a strongly ionophilic pore ($\delta E = -2.5 k_B T$) and for weakly ionophobic pores ($\delta E = 38.5 k_B T$ for $d=0.5$nm and $\delta E = 33.2k_BT$ for $d=0.6$nm). The insets show the 3D packing fraction $\eta_{3D} = (\pi/6) \rho d^3$ as a function of pore width.}
\label{fig:cap_pore_sim}
\end{figure*}


Figure~\ref{fig:capen_pore}b shows the energy per surface area stored for each of the two slit widths. For low applied voltages the stored energy is slightly higher for the weakly ionophobic pore (inset in Figure~\ref{fig:capen_pore}b) because it has a higher low-voltage capacitance. For intermediate voltages, however, capacitance is higher for ionophilic pores (\textit{c.f.}~Figure~\ref{fig:trans}a), and weak ionophobicity reduces the energy storage in this voltage range. For sufficiently high voltages, ionophilic pores eventually saturate, while ionophobic pores continue to charge, and this leads ultimately to higher energies stored by ionophobic pores~\cite{leePRL2014, kondrat:nh:16}.


Monte Carlo simulations confirm the predicted trends of the mean field model. Strongly ionophilic pores show a monotonic decrease of capacitance with increasing pore size (Figure~\ref{fig:cap_pore_sim}a), while weak ionophobicity produces a local maximum with capacitances comparable to, or even higher than, the capacitance at $L \approx d$ (Figure~\ref{fig:cap_pore_sim}b). The total ion packing fraction is almost constant with increasing slit width for strongly ionophilic pores (inset of Figure~\ref{fig:cap_pore_sim}a) whereas for ionophobic pores it decreases with increasing slit width (inset of Figure ~\ref{fig:cap_pore_sim}b), agreeing qualitatively with the mean field model (inset of Figure~\ref{fig:capen_pore}a). We note that a slightly smaller $\delta E$ in mean field theory is sufficient to achieve the same ion occupancy at zero voltage, hence ionophobicity, as in the Monte Carlo simulations. This discrepancy is due to the fact that our mean field theory does not account for the change in ion-image interactions due to ions positioning themselves off the central symmetry plane of the pore (though this effect does not change the qualitative predictions of the mean field model). 

\begin{figure*}
\centering
\includegraphics[width=1.0\textwidth]{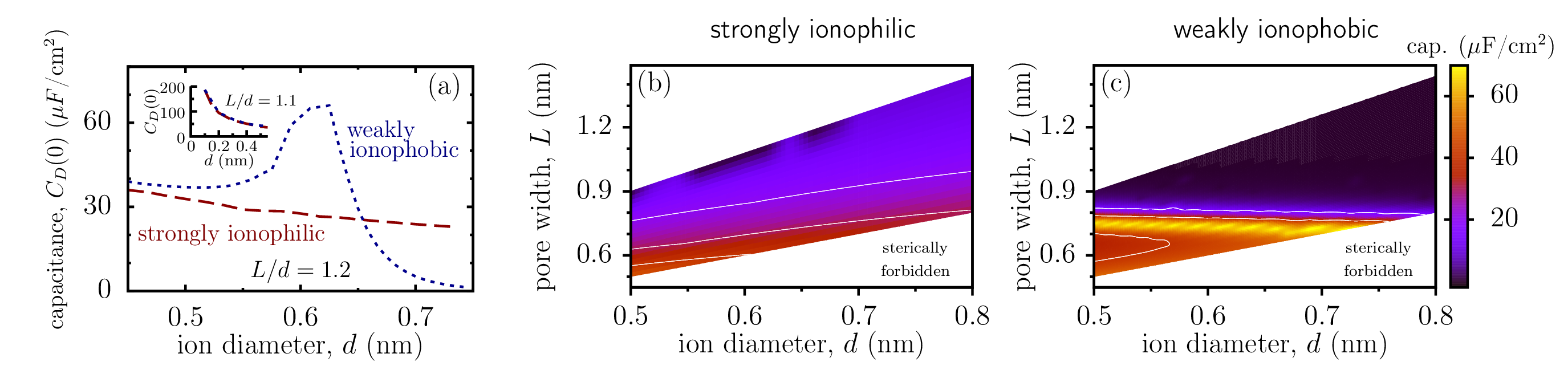} 
\caption{(a) Capacitance at zero voltage $C_D(0)$ as a function of ion diameter ($d=d_\pm$) for weakly ionophobic ($\delta E= 25k_BT$) and strongly ionophilic ($\delta E= 10k_BT$) pores calculated using mean field theory (MFT). The ratio between the pore width and the ion diameter is fixed to $L/d = 1.2$. The inset shows $C_D(0)$ for small ions with $d \lesssim 0.5$nm and $L/d=1.1$. For such ultra narrow pores, $C_D(0)$ is practically independent of ionophobicity (for the values of $\delta E$ used in this plot). This is because the image forces (see Eq.~(\ref{eq:self})) overcompensate $\delta E$ for slit widths $L \lesssim 0.55$, making the pores completely filled with ions. (b)-(c) Capacitance map for strongly and weakly ionophilic nanopores and room-temperature ionic liquids (with $d \gtrsim 0.5$nm) in the plane of ion diameter ($d$) and the pore width ($L$) calculated using MFT. This figure suggests a `two-step optimization strategy', in which an ionophobicity-dependent optimal \emph{pair} ($d,L$) exists that maximizes the differential capacitance.}
\label{fig:cap_ion}
\end{figure*} 


The presence of a local maximum is distinct from the oscillatory behaviour of capacitance as a function of pore width \cite{feng2011supercapacitor,jiang2011oscillation,pizio2012electric} and solvent polarity \cite{jiang:nanoscale:14} observed in systems with wider pores. For weakly ionophobic pores, a maximum occurs because of entropic effects, while in the cited works it is the overlapping double layers and highly polar solvent, respectively, that cause the non-monotonicity in the capacitance (notice a second maximum in Figure~\ref{fig:cap_pore_sim}a, with a similar origin but with a much smaller amplitude). 

Another important aspect of capacitance optimization is how to choose the ion size ($d$) and slit width ($L$). Intuitively, one might expect that it is the \emph{ratio} between the two that affects the capacitance. However, Figure~\ref{fig:cap_ion}a shows that for weakly ionophobic pores, a peak in the capacitance emerges as the ion diameter increases while the ratio between the slit width and the ion diameter, $L/d$, is kept fixed. This peak occurs because the ion self-energy and the metallic screening decrease (\textit{i.e.}~electrostatic interactions become stronger) when increasing the ion diameter at constant $L/d$ (see Eqs.~(\ref{interaction_potential}) and (\ref{eq:self}) in Appendix~\ref{app:mft}). This effect makes the pore less populated for weak ionophobicities, and thus adsorption of new counterions becomes entropically more favorable, giving rise to a local maximum at intermediate ion sizes. For strongly ionophilic pores, the ion density is close to maximal and charging proceeds mainly via swapping co-ions for counterions. Therefore, the capacitance increases monotonically with decreasing ion diameter (dashed curve in Figure~\ref{fig:cap_ion}a). 


From these results we see that the highest possible capacitance may be obtained by optimizing \emph{both} the pore width and the ion diameter (see Figures~\ref{fig:cap_ion}b and c). Crucially, the position of this optimum in parameter space, and its properties, depend on the pore's ionophilicity. For strongly ionophilic pores, the pores are completely filled with ions and a maximal capacitance is achieved for small ions and tight pores, \textit{i.e.}~$L_\mathrm{opt} \approx d$ for any $d$ and the capacitance increases as $d$ decreases (Figure~\ref{fig:cap_ion}a and b). Remarkably, a strong increase of capacitance is obtained for pores below $0.5$~nm, reaching the values as high as 200~$\mu$F/cm$^2$ (see the inset in Figure~\ref{fig:cap_ion}a). This result suggests that such ultranarrow pores combined with inorganic electrolytes with small ions may be beneficial for the charge and energy storage.





For weak ionophobicities, there is an optimal slit width $L_\mathrm{opt} \approx 0.74$nm that depends only weakly on the ion diameter $d$, for $d \gtrsim 0.5$nm (Figure~\ref{fig:cap_ion}c). As discussed, this is connected with emptying (half-filled) pores, which maximizes the capacitance, \textit{i.e.} $L_\mathrm{opt} \approx L_\mathrm{empty}$ (see Figures~\ref{fig:capen_pore}a and \ref{fig:cap_pore_sim}b). Since the image-force interactions are ion-size independent (see Eq.~(\ref{eq:self})), we roughly estimate that the pore is emptied when $L \gtrsim L_\mathrm{empty} \approx l_B \ln(2)/ \delta E$~\footnote{We note however that this reasoning is valid for pore-width independent (or only weakly dependent) resolvation energies. Although this is not generally the case, it seems reasonable to expect it to hold for small variations of the slit width $d < L < 2d$ considered in this work.}. For the parameters of Figure~\ref{fig:cap_ion}c, $l_B = 25$nm and $\delta E = 25$ (in units of $k_BT$), we obtain $L_\mathrm{empty} \approx 0.7$nm. This value is slightly modified by the ion-size dependent entropic and screened electrostatic interactions, which additionally bring a weak $L_\mathrm{opt} (d)$ dependence.


To conclude this section, the fact that the capacitance reaches a maximum for pores that are significantly wider than the ion diameter may have an important impact on optimizing supercapacitors. It has previously been assumed that there is necessarily a trade-off between having large capacitance (narrow pores) and fast charging (wide pores) \cite{kondrat:jpcc:13}. Charging is slower for narrower pores because the collective (or effective) diffusion coefficient, $D_{eff}$, which determines the rate of charging, decreases with decreasing the pore width, $L$ \cite{kondrat:jpcc:13, kondrat:nm:14, lee:nanotech:14}; for instance,  $D_{eff}$ is almost doubled when $L$ increases from $0.7$nm to $0.9$nm~\cite{kondrat:jpcc:13}. Our analysis therefore provides the key insight that charging kinetics and capacitance can be simultaneously optimized by tuning the ionophilicity of the pore. 

\section{Hysteretic charging}
\label{hyster_sect}

\begin{figure*}
\vspace{-1cm}
\centering
\includegraphics[width=0.7\textwidth]{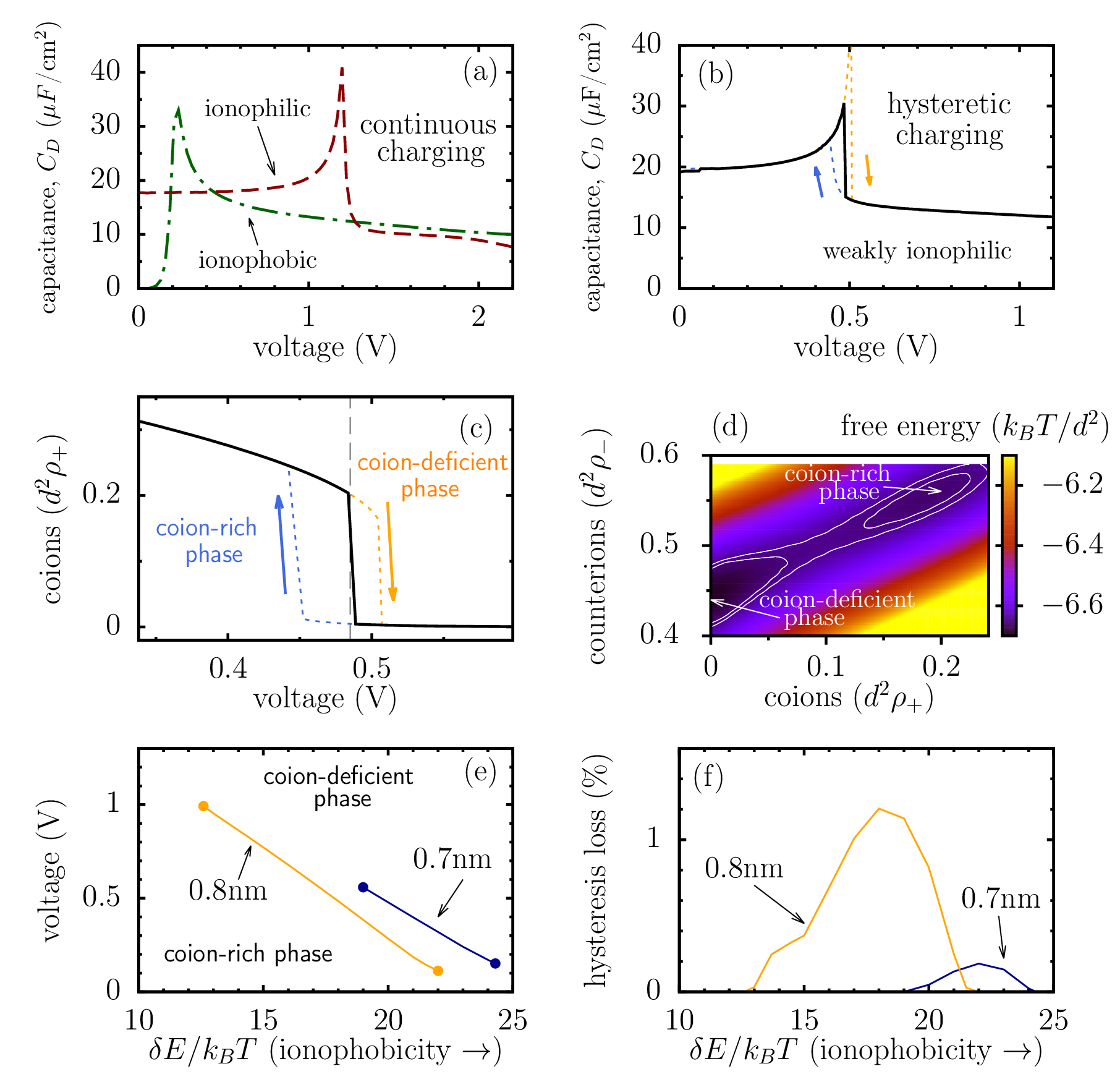}
\caption{Charging proceeds via a discontinuous phase transition for some parameter regimes. In all plots ion diameter $d=0.5$nm, and the mean-field theory is used to compute the thermodynamic properties. (a) Capacitance as a function of voltage in the continuous charging regime. The ionophilic  and ionophobic pores have resolvation energies $\delta E = 10k_BT$ and $\delta E = 32k_BT$, respectively, and width $L=0.8$nm. (b) Capacitance as a function of voltage in the discontinuous charging regime. The solid curve indicates the minimum free energy path, and the dotted curves denote the hysteresis loop with the arrows showing charging/discharging routes. The curve is plotted for $\delta E= 18k_BT$ and $L=0.8$nm. (c)~Discontinuous charging is connected with an abrupt expulsion of coions. We identify here coion-deficient and coion-rich phases with vanishing and appreciable amount of coions in the pore, respectively. The arrows show charging and discharging paths, which follow the metastable branches denoted by dotted lines, with the solid curve giving the minimum energy path, all as in panel (b). The vertical dashed line marks a voltage $V_\textrm{coex}\approx 0.49$~volts at which the two phases have the same free energy and coexist. (d) The free energy landscape for applied voltage $V_\textrm{coex}$ corresponding to the dash line in panel (c). White lines are a contour plot of constant free energy and the arrows point to the ion densities which simultaneously minimize the free energy at coexistence. The contour plot has been obtained by evaluating the free energy for given densities of cations and anions using Eq.~(\ref{eq:fe}). (e) The phase diagram plotted for different slit widths. (f) The percentage of energy that is lost due to hysteresis during discharging of a nanopore charged at 1.3 volts (see Appendix~\ref{app:hyster}). }
\label{fig:trans}
\end{figure*}

Next, we study the charging hysteresis for non-zero applied voltages. Figure \ref{fig:trans} shows that charging can proceed via a first-order discontinuous phase transition at intermediate ionophilicities, or via a continuous process beyond the critical endpoints.

For strongly ionophobic pores (with resolvation energies $\delta E$ above the critical endpoints, Figure~\ref{fig:trans}e), the pore is empty at no applied voltage, and counter-ions only enter the pore when the applied potential matches the unfavourable resolvation energy; hence there is a peak in capacitance when charging starts at $e V \approx \delta E$ (Figure~\ref{fig:trans}a). Subsequently a separate regime of packing like charges starts and persists until the applied potential surpasses the interaction energy between co-ions. Beyond that point the capacitance falls off rapidly. 


The situation is different for strongly ionophilic pores with $\delta E$ below the critical endpoints, where the pore is nearly fully occupied with ions at zero voltage. At low voltages, charging proceeds via swapping of co-ions for counter-ions and co-ion desorption, while at higher voltages (above $\approx 1.2$~V) it is the counter-ion adsorption that drives the charging. It has been shown elsewhere~\cite{kondrat:nh:16} that adsorption leads to low capacitances, as compared to swapping and desorption, and hence the drop in the differential capacitance, $C_D(V)$, at $V \approx 1.2$~V (see the dash line in Figure~\ref{fig:trans}a).

For ionophilicities between the two critical endpoints, charging proceeds via a large discontinuous drop in capacitance and ion densities (Figures~\ref{fig:trans}b and c respectively). At low voltages, charging is driven by swapping of co-ions for counter-ions and co-ion expulsion, but at the point of the phase transition the system expels all co-ions. This is similar to strongly ionophilic pores (Figure~\ref{fig:trans}a), except that this process is discontinuous and we can observe two coexisting phases~\cite{kondrat2011superionic}: Coion-deficient phase, with a vanishing number of coions in the pore, and coion-rich phase with the essentially non-zero coion density (Figures \ref{fig:trans}c-d). Within a small voltage window around this coexistence point, one of the two phases is metastable and there is a free-energy barrier between them (Figure~\ref{fig:trans}d); as we will see this may have an important implication for the energy storage. Beyond the metastability window charging occurs solely via counter-ion adsorption.

This voltage-induced abrupt expulsion of co-ions and the drop in the total ion density has recently been observed in atomistic molecular dynamics simulations of [C2mim][FSI] and [C4mim][TFSI] ionic liquids in ($0.75$nm wide) slit nanopores~\cite{vatamanu:acsnano:15}. There, however, the drop in coion density is smoother than we observe, which is most likely due to finite size effects. Indeed, in our theory the pore is formally infinite in the lateral directions, while in the simulations of Ref.~\cite{vatamanu:acsnano:15} the pore was only about a few tens nm long.

Our results are also in line with recent NMR experiments~\cite{griffin:fd:14, griffin:natmat:15} which show that charging proceeds continuously (\textit{i.e.}~without phase transitions or accompanying abrupt co-ion expulsion) for [PEt$_4$][BF$_4$] ionic liquids and KOH activated carbon pores, known to be strongly ionophilic~\cite{richey:jpcc:philicity:14}. While it is difficult to calculate the resolvation energy (ionophobicity) for this system, Figure~\ref{fig:trans}e suggests that it must at least be lower than $12 k_BT$. This is consistent with our estimate of the crossover between the ionophilic and ionophobic pores which occurs at $\delta E_\mathrm{crossover} \approx 17 k_BT$ for the $1$nm wide pore of Ref.~\cite{griffin:natmat:15} (see Section~\ref{sec:ionophobicity}).


Figure \ref{fig:trans}e shows that the transition voltage increases with decreasing ionophobicity, as both counter ions and co-ions are favourably adsorbed into the pore, making it more difficult to expel the co-ions. The window of ionophilicities for which a phase transition occurs is wider for larger pores where electrostatic interactions are stronger and more long-ranged. 

A signature of a discontinuous phase transition is hysteresis as the system follows the locally metastable branch. The dotted lines in Figures \ref{fig:trans}b-c show that the branch followed by the system when the voltage is increased is different to that followed by the system for decreasing voltage. Figure \ref{fig:trans}f shows that a significant amount of energy is lost per cycle as a consequence of hysteresis, with wider pores producing larger energy losses. In practical applications, the regime where charging proceeds via a first order phase transition should therefore be avoided, but this hysteresis loop can be used to probe the properties of the supercapacitor system experimentally. It is likely however that these transitions will be smoothed out by the distribution of pore sizes~\cite{kondrat:ees:12} in typically used porous electrodes (for instance in popular carbide-derived carbons~\cite{simon2008materials}), making it difficult to capture them directly by \textit{in situ} NMR spectroscopy~\cite{wang:jacs:NMRsupercap:13, forse:pccp:13} or in electrochemical quartz crystal microbalance experiments~\cite{levi:nm:eqcm:09, levi:jpcc:13:EQCM, tsai:jacs:14}. However, novel graphene-based nanoporous electrodes~\cite{yoo:nanolett:graphene:11, yang:sci13:graphene}, which consist of nearly unimodal well-aligned slit pores with controllable pore widths, seem a promising candidate for validating our predictions experimentally. Although the quantum capacitance of these materials, which we have neglected in this work, may change the values of the total capacitance~\cite{tao:natnano:graphene:09, stoller:ees:graphene:11, kornyshev:jsse:2double:14}, we do not expect that it will affect the existence of phase transitions or their order, as they result solely from the competition between entropic effects (due to confinement) and screened inter-ionic interactions induced by the conducting pore walls. Nevertheless it would be very interesting to analyze the role that the quantum capacitance may play in energy storage and hysteretic charging.



\section{Conclusion}

Using a mean-field model for charge storage in 2D nano-confinement and Monte Carlo simulations of the corresponding system in 3D, we have demonstrated the possibility of simultaneously boosting capacitance, accelerating charging, and avoiding hysteresis in nanoporous supercapacitors. Our calculations show that high capacitances can be achieved for electrolytes with small ions and close-fitting pores (inset in Figure~\ref{fig:cap_ion}a). On the other hand, we find that the long-espoused paradigm that equates narrow pores with necessarily larger capacitance does not always hold: Entropic effects may produce a second, more pronounced peak in capacitance for relatively wide ionophobic pores and room-temperature ionic liquids (Fig.~\ref{fig:capen_pore}a and \ref{fig:cap_pore_sim}b). Similarly, there exists an optimal ion diameter when the ratio between the ion diameter ($d$) and the pore width  ($L$) is kept constant, suggesting an optimization of supercapacitors with respect to the $(d,L)$ pair rather than the pore size alone (Fig.~\ref{fig:cap_ion}).

For non-zero applied voltages, a voltage-induced discontinuous phase transition is predicted by the model, and the phase diagram has two critical endpoints corresponding to the limit of very ionophobic and ionophilic pores (Fig.~\ref{fig:trans}). The phase transition gives rise to hysteresis and a sizeable energy loss, but crucially can be avoided by either reducing the operating voltage of the capacitor, or by pushing the pore ionophilicity away from the critical endpoints (for instance by making pores more ionophobic). Thus, the capacitance-power-hysteresis trilemma can be resolved by judiciously tuning material parameters and operating ranges.

Ion diameter and pore width are relatively straightforward to tune experimentally; our framework shows how to optimize both parameters together. We also single out a key outstanding challenge for further experimentation --- the controlled tuning of the ionophobicity/ionophilicity of the pore. The ionophilicity can be controlled by changing the ion-pore non-electrostatic interaction. In addition to van der Waals interactions, suppose we have an ion that has multiple conformational states with different effective diameters (for example expanded versus folded alkyl chains on an ionic liquid ion). If the pore separation is less than the effective diameter of the lowest energy conformation, then entering the pore will require the ion to adopt a higher conformation energy, and thus incur a conformational energy penalty. Similarly, if the pore walls are flexible and ions can only enter by deforming the walls (as observed in Refs.~\cite{kaasik2013anisometric,hantel2014parameters}), the elastic energy will result in an unfavourable ion-pore non-electrostatic interaction. This interaction could also be controlled by addition of surfactants~\cite{fic:ea:10, fic:ea:11} or solvents~\cite{griffin:natmat:15, force:jacs:NMRDyn, rochester:1d}, or by using ionic liquid  mixtures~\cite{lin:jpcl:11}, or via functionalisation of porous carbons~\cite{richey:jpcc:philicity:14}. Experimental studies of ionophilicity are currently scarce, however, and we hope that our theory will provide a framework to assess and direct future efforts to address this.  

\begin{acknowledgments}

This work is supported by an EPSRC Doctoral Training Award and Fulbright Scholarship to AAL. SK~acknowledges COST Action MP1004 for supporting his short-time scientific visit to Imperial College London where part of this work has been done, and appreciates fruitful discussions with Alexei Kornyshev (Imperial College).

\end{acknowledgments}

\appendix
\section{Improved mean-field approximation}
\label{app:mft}

The Helmholtz free energy $F(\rho_+,\rho_-)$ per unit area of a homogeneous ionic liquid in a narrow slit nanopore is given by
\begin{equation}
\beta F = U_{\rm el}(\rho_+,\rho_-)  - S(\rho_+,\rho_-) + \sum_{\alpha = \pm} h_\alpha \rho_\alpha, 
\label{eq:free_energy_fun}
\end{equation}
where $\beta=(k_BT)^{-1}$ ($k_B$ is the Boltzmann constant and $T$ temperature), $\rho_{\pm}$ are the two-dimensional ion densities, and it is convenient to introduce $\rho = \rho_+ + \rho_-$ and $c = \rho_+ - \rho_-$. $k_BTU_{\rm el}$ is the contribution to free energy due to electrostatic interactions, $k_BTS$ is the excluded volume entropic contribution, and  $k_BTh_\alpha$ is the electrochemical potential of the ion species (see below). 

\subsection{Electrostatic interactions}
The electrostatic interaction can be expressed in terms of direct correlation functions $\mathcal{C}_{\alpha\beta}(x,x')$ using a functional Taylor expansion 
\begin{equation}
U_{\rm el}(\rho_+,\rho_-) \approx \sum_{\alpha,\beta =\pm} \int \mathrm{d}x \int \mathrm{d}x'
	 \mathcal{C}_{\alpha \beta}(x,x') \rho_{\alpha} \rho_{\beta} ,  
\end{equation}
with $\mathcal{C}_{\alpha\beta}(x,x')$ computed using the Ornstein-Zernicke relation with an appropriate closure \cite{Hansen2014}.  However, we take here a simplifying assumption in the spirit of the random phase approximation (RPA). In RPA, the direct correlation function is approximated by the interaction potential, which is valid for asymptotically weak interactions but overestimates the correlations at close particle separations \cite{Hansen2014}. To improve on this, we assume that the direct correlation function is essentially negligible for distances less than the average particle separation; beyond this distance the interaction is weak and can be approximated by the classical RPA. As such,  $\mathcal{C}_{\alpha \beta} (x,x') \approx v_{\alpha\beta}(x-x') \theta(|x-x'|-R_{\mathrm{c}})$ where $v_{\alpha \beta}(x-x')$ is the electrostatic interaction kernel and $R_{\mathrm{c}}=1/\sqrt{\pi \rho}$ is the average separation between particles (this result is also known as the cut-out disc approximation, see Ref.~\cite{kornyshev:jeac:86}). 

The interaction potential between two point charges confined in a slit metallic nanopore and separated by distance $r$ is given by~\cite{smythe:book}
\begin{align}
	v_{\alpha\beta}(r, z_1,z_2) =  \frac{4 q_\alpha q_\beta}{\varepsilon_p L} 
		\sum_{n=1}^{\infty} K_0\left(\pi n r/L \right)
		\sin(\pi n z_1/L)\sin(\pi n z_2/L),
\label{interaction_potential}
\end{align} 
where $q_\alpha$ and $q_\beta$ are charges, $z_1$ and $z_2$ are ion positions across the pore (within the mean-field model we assume $z_{1,2}=L/2$, but we use the full potential in our Monte Carlo simulations, see below), and $\varepsilon_p$ is the dielectric constant in the pore; we have taken $\varepsilon_p = 2$ in all calculations, but we note that $\varepsilon_p$ shall in principle depend on the pore width (and voltage), and this dependence may have a profound effect on system's behaviour~\cite{kondrat:ec:13}. 

For monovalent ions we obtain~\cite{kondrat2011superionic} 
\begin{equation}
U_{\rm el}(c, \rho) = 4c^2 R_c(\rho) l_B  \sum_{n=1}^{\infty} \frac{\sin^2(\pi n /2)}{n} K_1\left(\pi n R_c(\rho) /L\right),
\end{equation}
where $l_B = e^2/(\varepsilon_p k_B T)$ is the thermal Bjerrum length. Constrained ionic motion within the pore means that effectively only electronic degrees of freedom contribute to dielectric screening and we take $\epsilon_p = 2$, the high frequency dielectric constant of typical ionic liquids. It is easy to see that at constant $L/d$, the functional dependence of the potential $v_{\alpha\beta}$ on $d$ is the same as on $L$ at constant $d$.

\subsection{Entropic contributions}
The entropic contribution in Eq.~(\ref{eq:free_energy_fun}) is modelled here based on the analytically solvable 2D scaled particle theory. The typical pore separation is only slightly larger than the ion diameter, and thus we can take into account the positional disorder of ions normal to the pore surface as a perturbation to the otherwise 2D system --- this effectively renormalizes the ion diameter. Thus the entropy is given as a sum of the ideal gas contribution and hard core exclusion,   
\begin{equation}
S(\rho_{+},\rho_{-}) = \sum_{\alpha = \pm} \tilde \rho_{\alpha}\ln \tilde \rho_\alpha 
	+ \tilde \rho \left[ \frac{\tilde \eta^2}{ 1-\tilde \eta} - \ln (1-\tilde \eta) \right],
\end{equation}
where $\tilde{\eta}(\rho) = \pi \rho \sigma^2/4 = \pi \tilde \rho d^2/4$ is the effective ion packing fraction and $\sigma(d,L,\rho)$ is a renormalized diameter that accounts for the out-of-plane packing of ions. \citeauthor{schmidt1996freezing}~\cite{schmidt1996freezing,schmidt1997phase} showed that 
\begin{equation}
\sigma^2(d,L,\rho) = d^2 + \frac{1}{\alpha(\rho)}  - \frac{L \exp\left(\alpha(\rho) L^2/4\right) }{\sqrt{\pi \alpha(\rho) } \mathrm{erfi}(\sqrt{\alpha(\rho)} L/2)}
\end{equation}  
where $\mathrm{erfi}(z)$ is the imaginary error function, and 
\begin{equation}
\alpha(\rho) = \pi \rho g(\rho) = \pi \rho \frac{1-\tilde{\eta}(\rho)/2}{(1- \tilde{\eta}(\rho))^2}
\end{equation} 
is the average density at contact and $g(\rho)$ the 2D pair correlation function evaluated at contact.  


\subsection{Electrochemical potential}
\label{app:mft:pot}

The electrochemical potential (in infinite dilution) is given by
\begin{align}
h_{\pm} = \pm u + \delta E_{\mathrm{self}} + \delta E_{\pm} =  \pm u - \frac{l_B }{L}\ln 2  + \delta E_{\pm}.
\label{chemical_pot}
\end{align} 
Here the first term results from the applied voltage, and the second term originates from the ion self energy
\begin{align}
\label{eq:self}
	\delta E_{\mathrm{self}}(z)  &= \lim_{r\rightarrow 0} \left(\phi(r) - \frac{1}{r} \right) 
	=  \frac{l_B}{L} \int_0^\infty\left[\frac{\sinh(Q(1-z/L))\sinh(Qz/L)}{\sinh(Q)}\right.
	\left. ~ - \frac{1}{2} \right]dQ,
\end{align}
where $z$ is the position across the pore and $r$ distance to the charge (see Ref.~\cite{kondrat2011superionic}). At the pore mid plane $z=L/2$ which gives $\delta E_{\mathrm{self}}(L/2) = - l_B \ln(2)/L$. Again, $\delta E_{\mathrm{self}}$ depends on $L$ in the same way as on $d$ at constant $L/d$. The last term in Equation (\ref{chemical_pot}), $\delta E_{\pm}$,  is the ``resolvation energy'' which is the energy of transferring an ion from the bulk of a supercapacitor into the pore in the absence of other ions.


\subsection{Stored energy and hysteretic energy loss}
\label{app:hyster}

The energy (per surface area) stored in a nanopore by charging it from $u=u_1$ to $u=u_2$ is
\begin{align}
	\mathcal{E}(u_1,u_2) = \int_{u_1}^{u_2}  u \; C_D(u)   \; \mathrm{d}u,
\end{align}
where $C_D(u) = \mathrm{d}Q /\mathrm{d}V$ is differential surface-specific capacitance. To obtain an energy lost due to hysteresis (Figure~\ref{fig:trans}f), we first calculated the energy stored when charging a nanopore from $u_1=0$ to $u_2=V=1.3$~V along the charging path (orange lines/arrows in Figure~\ref{fig:trans}b), $\mathcal{E}_\textrm{charg} = \mathcal{E}(0,V)$; and the energy released by fully discharging it along the discharging path (blue lines/arrows in Figure~\ref{fig:trans}b), $\mathcal{E}_\textrm{discharg} = - \mathcal{E}(V,0)$. The hysteretic energy loss is then shown as percentage of the stored energy lost in hysteresis, i.e.~$\boldsymbol{(} \mathcal{E}_\mathrm{charg} - \mathcal{E}_\mathrm{discharg}\boldsymbol{)}/ \mathcal{E}_\mathrm{charg}$.

\section{Grand canonical Monte Carlo simulations}
\label{app:mc}

We used the same method as in Ref.~\cite{kondrat2011superionicMC}, so we will only summarize the simulation method here, and refer the reader to~\cite{kondrat2011superionicMC} for further detail.

Ionic liquid molecules are modelled as charged hard spheres, but instead of the Coulomb potential, we use the analytical solution (\ref{interaction_potential}) for the ion-ion interactions. Ion-pore wall interactions are captured by potential (\ref{eq:self}), which accounts for image forces. The resolvation energy ($\delta E$) and the applied voltage ($V$) are subsumed into the chemical potential in the grand canonical simulations, $\mu_\pm^{(sim)} = \delta E \pm e V$, where $e$ is the elementary charge, and $\delta E$ is the resolvation energy, as before. In all our simulations we took temperature $T=328$K and the relative dielectric constant inside pores $\varepsilon_p = 2$.

To calculate differential capacitance, we differentiated the accumulated charge obtained from Monte Carlo simulations with respect to voltage numerically using the Holoborodko method with seven points \cite{holoborodko:snrd}.

\bibliography{supercap_hysteresis}

\begin{thebibliography}{73}%
\makeatletter
\providecommand \@ifxundefined [1]{%
 \@ifx{#1\undefined}
}%
\providecommand \@ifnum [1]{%
 \ifnum #1\expandafter \@firstoftwo
 \else \expandafter \@secondoftwo
 \fi
}%
\providecommand \@ifx [1]{%
 \ifx #1\expandafter \@firstoftwo
 \else \expandafter \@secondoftwo
 \fi
}%
\providecommand \natexlab [1]{#1}%
\providecommand \enquote  [1]{``#1''}%
\providecommand \bibnamefont  [1]{#1}%
\providecommand \bibfnamefont [1]{#1}%
\providecommand \citenamefont [1]{#1}%
\providecommand \href@noop [0]{\@secondoftwo}%
\providecommand \href [0]{\begingroup \@sanitize@url \@href}%
\providecommand \@href[1]{\@@startlink{#1}\@@href}%
\providecommand \@@href[1]{\endgroup#1\@@endlink}%
\providecommand \@sanitize@url [0]{\catcode `\\12\catcode `\$12\catcode
  `\&12\catcode `\#12\catcode `\^12\catcode `\_12\catcode `\%12\relax}%
\providecommand \@@startlink[1]{}%
\providecommand \@@endlink[0]{}%
\providecommand \url  [0]{\begingroup\@sanitize@url \@url }%
\providecommand \@url [1]{\endgroup\@href {#1}{\urlprefix }}%
\providecommand \urlprefix  [0]{URL }%
\providecommand \Eprint [0]{\href }%
\providecommand \doibase [0]{http://dx.doi.org/}%
\providecommand \selectlanguage [0]{\@gobble}%
\providecommand \bibinfo  [0]{\@secondoftwo}%
\providecommand \bibfield  [0]{\@secondoftwo}%
\providecommand \translation [1]{[#1]}%
\providecommand \BibitemOpen [0]{}%
\providecommand \bibitemStop [0]{}%
\providecommand \bibitemNoStop [0]{.\EOS\space}%
\providecommand \EOS [0]{\spacefactor3000\relax}%
\providecommand \BibitemShut  [1]{\csname bibitem#1\endcsname}%
\let\auto@bib@innerbib\@empty
\bibitem [{\citenamefont {Lu}\ \emph {et~al.}(2013)\citenamefont {Lu},
  \citenamefont {Beguin},\ and\ \citenamefont
  {Frackowiak}}]{lu2013supercapacitors}%
  \BibitemOpen
  \bibfield  {author} {\bibinfo {author} {\bibfnamefont {Max}\ \bibnamefont
  {Lu}}, \bibinfo {author} {\bibfnamefont {Francois}\ \bibnamefont {Beguin}}, \
  and\ \bibinfo {author} {\bibfnamefont {Elzbieta}\ \bibnamefont
  {Frackowiak}},\ }\href@noop {} {\emph {\bibinfo {title} {Supercapacitors:
  Materials, Systems and Applications}}}\ (\bibinfo  {publisher} {John Wiley \&
  Sons},\ \bibinfo {year} {2013})\BibitemShut {NoStop}%
\bibitem [{\citenamefont {Simon}\ and\ \citenamefont
  {Gogotsi}(2013)}]{simon2012capacitive}%
  \BibitemOpen
  \bibfield  {author} {\bibinfo {author} {\bibfnamefont {P.}~\bibnamefont
  {Simon}}\ and\ \bibinfo {author} {\bibfnamefont {Y.}~\bibnamefont
  {Gogotsi}},\ }\bibfield  {title} {\enquote {\bibinfo {title} {Capacitive
  energy storage in nanostructured carbon{--}electrolyte systems},}\
  }\href@noop {} {\bibfield  {journal} {\bibinfo  {journal} {Acc. Chem. Res.}\
  }\textbf {\bibinfo {volume} {46}},\ \bibinfo {pages} {1094} (\bibinfo {year}
  {2013})}\BibitemShut {NoStop}%
\bibitem [{\citenamefont {Simon}\ and\ \citenamefont
  {Gogotsi}(2008)}]{simon2008materials}%
  \BibitemOpen
  \bibfield  {author} {\bibinfo {author} {\bibfnamefont {P.}~\bibnamefont
  {Simon}}\ and\ \bibinfo {author} {\bibfnamefont {Y.}~\bibnamefont
  {Gogotsi}},\ }\bibfield  {title} {\enquote {\bibinfo {title} {Materials for
  electrochemical capacitors},}\ }\href@noop {} {\bibfield  {journal} {\bibinfo
   {journal} {Nature Mater.}\ }\textbf {\bibinfo {volume} {7}},\ \bibinfo
  {pages} {845--854} (\bibinfo {year} {2008})}\BibitemShut {NoStop}%
\bibitem [{\citenamefont {Frackowiak}(2007)}]{frackowiak:07}%
  \BibitemOpen
  \bibfield  {author} {\bibinfo {author} {\bibfnamefont {E.}~\bibnamefont
  {Frackowiak}},\ }\bibfield  {title} {\enquote {\bibinfo {title} {Carbon
  materials for supercapacitor application},}\ }\href@noop {} {\bibfield
  {journal} {\bibinfo  {journal} {Phys. Chem. Chem. Phys.}\ }\textbf {\bibinfo
  {volume} {9}},\ \bibinfo {pages} {1774} (\bibinfo {year} {2007})}\BibitemShut
  {NoStop}%
\bibitem [{\citenamefont {Gogotsi}\ \emph {et~al.}(2003)\citenamefont
  {Gogotsi}, \citenamefont {Nikitin}, \citenamefont {Ye}, \citenamefont {Zhou},
  \citenamefont {Fischer}, \citenamefont {Yi}, \citenamefont {Foley},\ and\
  \citenamefont {Barsoum}}]{gogotsi:natmat:03}%
  \BibitemOpen
  \bibfield  {author} {\bibinfo {author} {\bibfnamefont {Yury}\ \bibnamefont
  {Gogotsi}}, \bibinfo {author} {\bibfnamefont {Alexei}\ \bibnamefont
  {Nikitin}}, \bibinfo {author} {\bibfnamefont {Haihui}\ \bibnamefont {Ye}},
  \bibinfo {author} {\bibfnamefont {Wei}\ \bibnamefont {Zhou}}, \bibinfo
  {author} {\bibfnamefont {John~E.}\ \bibnamefont {Fischer}}, \bibinfo {author}
  {\bibfnamefont {Bo}~\bibnamefont {Yi}}, \bibinfo {author} {\bibfnamefont
  {Henry~C.}\ \bibnamefont {Foley}}, \ and\ \bibinfo {author} {\bibfnamefont
  {Michel~W.}\ \bibnamefont {Barsoum}},\ }\bibfield  {title} {\enquote
  {\bibinfo {title} {Nanoporous carbide-derived carbon with tunable pore
  size},}\ }\href {\doibase 10.1038/nmat957} {\bibfield  {journal} {\bibinfo
  {journal} {Nat. Mater.}\ }\textbf {\bibinfo {volume} {2}},\ \bibinfo {pages}
  {591} (\bibinfo {year} {2003})}\BibitemShut {NoStop}%
\bibitem [{\citenamefont {Yoo}\ \emph {et~al.}(2011)\citenamefont {Yoo},
  \citenamefont {Balakrishnan}, \citenamefont {Huang}, \citenamefont {Meunier},
  \citenamefont {Sumpter}, \citenamefont {Srivastava}, \citenamefont {Conway},
  \citenamefont {Reddy}, \citenamefont {Yu}, \citenamefont {Vajtai},\ and\
  \citenamefont {Ajayan}}]{yoo:nanolett:graphene:11}%
  \BibitemOpen
  \bibfield  {author} {\bibinfo {author} {\bibfnamefont {J.~J.}\ \bibnamefont
  {Yoo}}, \bibinfo {author} {\bibfnamefont {K.}~\bibnamefont {Balakrishnan}},
  \bibinfo {author} {\bibfnamefont {J.~S.}\ \bibnamefont {Huang}}, \bibinfo
  {author} {\bibfnamefont {V.}~\bibnamefont {Meunier}}, \bibinfo {author}
  {\bibfnamefont {B.~G.}\ \bibnamefont {Sumpter}}, \bibinfo {author}
  {\bibfnamefont {A.}~\bibnamefont {Srivastava}}, \bibinfo {author}
  {\bibfnamefont {M.}~\bibnamefont {Conway}}, \bibinfo {author} {\bibfnamefont
  {A.~L.~M.}\ \bibnamefont {Reddy}}, \bibinfo {author} {\bibfnamefont
  {J.}~\bibnamefont {Yu}}, \bibinfo {author} {\bibfnamefont {R.}~\bibnamefont
  {Vajtai}}, \ and\ \bibinfo {author} {\bibfnamefont {P.~M.}\ \bibnamefont
  {Ajayan}},\ }\bibfield  {title} {\enquote {\bibinfo {title} {Ultrathin planar
  graphene supercapacitors},}\ }\href@noop {} {\bibfield  {journal} {\bibinfo
  {journal} {Nano Lett.}\ }\textbf {\bibinfo {volume} {11}},\ \bibinfo {pages}
  {1423--1427} (\bibinfo {year} {2011})}\BibitemShut {NoStop}%
\bibitem [{\citenamefont {Yang}\ \emph {et~al.}(2013)\citenamefont {Yang},
  \citenamefont {Cheng}, \citenamefont {Wang}, \citenamefont {Qiu},\ and\
  \citenamefont {Li}}]{yang:sci13:graphene}%
  \BibitemOpen
  \bibfield  {author} {\bibinfo {author} {\bibfnamefont {Xiaowei}\ \bibnamefont
  {Yang}}, \bibinfo {author} {\bibfnamefont {Chi}\ \bibnamefont {Cheng}},
  \bibinfo {author} {\bibfnamefont {Yufei}\ \bibnamefont {Wang}}, \bibinfo
  {author} {\bibfnamefont {Ling}\ \bibnamefont {Qiu}}, \ and\ \bibinfo {author}
  {\bibfnamefont {Dan}\ \bibnamefont {Li}},\ }\bibfield  {title} {\enquote
  {\bibinfo {title} {Liquid-mediated dense integration of graphene materials
  for compact capacitive energy storage},}\ }\href@noop {} {\bibfield
  {journal} {\bibinfo  {journal} {Science}\ }\textbf {\bibinfo {volume}
  {341}},\ \bibinfo {pages} {534--537} (\bibinfo {year} {2013})}\BibitemShut
  {NoStop}%
\bibitem [{\citenamefont {Richey}\ \emph {et~al.}(2014)\citenamefont {Richey},
  \citenamefont {Tran}, \citenamefont {Kalra}, \citenamefont {Elabd},\ and\
  \citenamefont {A.}}]{richey:jpcc:philicity:14}%
  \BibitemOpen
  \bibfield  {author} {\bibinfo {author} {\bibfnamefont {Francis~W.}\
  \bibnamefont {Richey}}, \bibinfo {author} {\bibfnamefont {Chau}\ \bibnamefont
  {Tran}}, \bibinfo {author} {\bibfnamefont {Vibha}\ \bibnamefont {Kalra}},
  \bibinfo {author} {\bibnamefont {Elabd}}, \ and\ \bibinfo {author}
  {\bibfnamefont {Yossef}\ \bibnamefont {A.}},\ }\bibfield  {title} {\enquote
  {\bibinfo {title} {Ionic liquid dynamics in nanoporous carbon nanofibers in
  supercapacitors measured with in operando infrared
  spectroelectrochemistry},}\ }\href {\doibase 10.1021/jp506903m} {\bibfield
  {journal} {\bibinfo  {journal} {J. Phys. Chem. C}\ }\textbf {\bibinfo
  {volume} {118}},\ \bibinfo {pages} {21846} (\bibinfo {year}
  {2014})}\BibitemShut {NoStop}%
\bibitem [{\citenamefont {Conway}(1999)}]{conway1999electrochemical}%
  \BibitemOpen
  \bibfield  {author} {\bibinfo {author} {\bibfnamefont {B.~E.}\ \bibnamefont
  {Conway}},\ }\href@noop {} {\emph {\bibinfo {title} {Electrochemical
  supercapacitors: scientific fundamentals and technological applications}}}\
  (\bibinfo  {publisher} {Springer},\ \bibinfo {year} {1999})\BibitemShut
  {NoStop}%
\bibitem [{\citenamefont {Raymundo-Pi{\~{n}}ero}\ \emph
  {et~al.}(2006)\citenamefont {Raymundo-Pi{\~{n}}ero}, \citenamefont
  {Kierczek}, \citenamefont {Machnikowski},\ and\ \citenamefont
  {B{\'{e}}guin}}]{pinero:carbon:06}%
  \BibitemOpen
  \bibfield  {author} {\bibinfo {author} {\bibfnamefont {E.}~\bibnamefont
  {Raymundo-Pi{\~{n}}ero}}, \bibinfo {author} {\bibfnamefont {K.}~\bibnamefont
  {Kierczek}}, \bibinfo {author} {\bibfnamefont {J.}~\bibnamefont
  {Machnikowski}}, \ and\ \bibinfo {author} {\bibfnamefont {F.}~\bibnamefont
  {B{\'{e}}guin}},\ }\bibfield  {title} {\enquote {\bibinfo {title}
  {Relationship between the nanoporous texture of activated carbons and their
  capacitance properties in different electrolytes},}\ }\href@noop {}
  {\bibfield  {journal} {\bibinfo  {journal} {Carbon}\ }\textbf {\bibinfo
  {volume} {44}},\ \bibinfo {pages} {2498--2507} (\bibinfo {year}
  {2006})}\BibitemShut {NoStop}%
\bibitem [{\citenamefont {Chmiola}\ \emph {et~al.}(2006)\citenamefont
  {Chmiola}, \citenamefont {Yushin}, \citenamefont {Gogotsi}, \citenamefont
  {Portet}, \citenamefont {Simon},\ and\ \citenamefont
  {Taberna}}]{chmiola2006anomalous}%
  \BibitemOpen
  \bibfield  {author} {\bibinfo {author} {\bibfnamefont {J.}~\bibnamefont
  {Chmiola}}, \bibinfo {author} {\bibfnamefont {G.}~\bibnamefont {Yushin}},
  \bibinfo {author} {\bibfnamefont {Y.}~\bibnamefont {Gogotsi}}, \bibinfo
  {author} {\bibfnamefont {C.}~\bibnamefont {Portet}}, \bibinfo {author}
  {\bibfnamefont {P.}~\bibnamefont {Simon}}, \ and\ \bibinfo {author}
  {\bibfnamefont {P.~L.}\ \bibnamefont {Taberna}},\ }\bibfield  {title}
  {\enquote {\bibinfo {title} {Anomalous increase in carbon capacitance at pore
  sizes less than 1 nanometer},}\ }\href@noop {} {\bibfield  {journal}
  {\bibinfo  {journal} {Science}\ }\textbf {\bibinfo {volume} {313}},\ \bibinfo
  {pages} {1760--1763} (\bibinfo {year} {2006})}\BibitemShut {NoStop}%
\bibitem [{\citenamefont {Largeot}\ \emph {et~al.}(2008)\citenamefont
  {Largeot}, \citenamefont {Portet}, \citenamefont {Chmiola}, \citenamefont
  {Taberna}, \citenamefont {Gogotsi},\ and\ \citenamefont
  {Simon}}]{largeot2008relation}%
  \BibitemOpen
  \bibfield  {author} {\bibinfo {author} {\bibfnamefont {C.}~\bibnamefont
  {Largeot}}, \bibinfo {author} {\bibfnamefont {C.}~\bibnamefont {Portet}},
  \bibinfo {author} {\bibfnamefont {J.}~\bibnamefont {Chmiola}}, \bibinfo
  {author} {\bibfnamefont {P.~L.}\ \bibnamefont {Taberna}}, \bibinfo {author}
  {\bibfnamefont {Y.}~\bibnamefont {Gogotsi}}, \ and\ \bibinfo {author}
  {\bibfnamefont {P.}~\bibnamefont {Simon}},\ }\bibfield  {title} {\enquote
  {\bibinfo {title} {Relation between the ion size and pore size for an
  electric double-layer capacitor},}\ }\href@noop {} {\bibfield  {journal}
  {\bibinfo  {journal} {J. Am. Chem. Soc.}\ }\textbf {\bibinfo {volume}
  {130}},\ \bibinfo {pages} {2730--2731} (\bibinfo {year} {2008})}\BibitemShut
  {NoStop}%
\bibitem [{\citenamefont {Lin}\ \emph {et~al.}(2009)\citenamefont {Lin},
  \citenamefont {Huang}, \citenamefont {Segalini}, \citenamefont {Largeot},
  \citenamefont {Taberna}, \citenamefont {Chmiola}, \citenamefont {Gogotsi},\
  and\ \citenamefont {Simon}}]{lin2009solvent}%
  \BibitemOpen
  \bibfield  {author} {\bibinfo {author} {\bibfnamefont {R.}~\bibnamefont
  {Lin}}, \bibinfo {author} {\bibfnamefont {P.}~\bibnamefont {Huang}}, \bibinfo
  {author} {\bibfnamefont {J.}~\bibnamefont {Segalini}}, \bibinfo {author}
  {\bibfnamefont {C{\'{e}}line}\ \bibnamefont {Largeot}}, \bibinfo {author}
  {\bibfnamefont {Pierre-Louis}\ \bibnamefont {Taberna}}, \bibinfo {author}
  {\bibfnamefont {John}\ \bibnamefont {Chmiola}}, \bibinfo {author}
  {\bibfnamefont {Yury}\ \bibnamefont {Gogotsi}}, \ and\ \bibinfo {author}
  {\bibfnamefont {Patrice}\ \bibnamefont {Simon}},\ }\bibfield  {title}
  {\enquote {\bibinfo {title} {Solvent effect on the ion adsorption from ionic
  liquid electrolyte into sub-nanometer carbon pores},}\ }\href@noop {}
  {\bibfield  {journal} {\bibinfo  {journal} {Electrochim. Acta}\ }\textbf
  {\bibinfo {volume} {54}},\ \bibinfo {pages} {7025--7032} (\bibinfo {year}
  {2009})}\BibitemShut {NoStop}%
\bibitem [{\citenamefont {Kondrat}\ and\ \citenamefont
  {Kornyshev}(2011)}]{kondrat2011superionic}%
  \BibitemOpen
  \bibfield  {author} {\bibinfo {author} {\bibfnamefont {S.}~\bibnamefont
  {Kondrat}}\ and\ \bibinfo {author} {\bibfnamefont {A.}~\bibnamefont
  {Kornyshev}},\ }\bibfield  {title} {\enquote {\bibinfo {title} {Superionic
  state in double-layer capacitors with nanoporous electrodes},}\ }\href@noop
  {} {\bibfield  {journal} {\bibinfo  {journal} {J. Phys.: Condens. Matter}\
  }\textbf {\bibinfo {volume} {23}},\ \bibinfo {pages} {022201} (\bibinfo
  {year} {2011})}\BibitemShut {NoStop}%
\bibitem [{\citenamefont {Rochester}\ \emph {et~al.}(2013)\citenamefont
  {Rochester}, \citenamefont {Lee}, \citenamefont {Pruessner},\ and\
  \citenamefont {Kornyshev}}]{Rochester2013}%
  \BibitemOpen
  \bibfield  {author} {\bibinfo {author} {\bibfnamefont {C.~Chris}\
  \bibnamefont {Rochester}}, \bibinfo {author} {\bibfnamefont {A.~Alpha}\
  \bibnamefont {Lee}}, \bibinfo {author} {\bibfnamefont {G.}~\bibnamefont
  {Pruessner}}, \ and\ \bibinfo {author} {\bibfnamefont {Alexei~A.}\
  \bibnamefont {Kornyshev}},\ }\bibfield  {title} {\enquote {\bibinfo {title}
  {Interionic interactions in electronically conducting confinement},}\
  }\href@noop {} {\bibfield  {journal} {\bibinfo  {journal} {ChemPhysChem}\
  }\textbf {\bibinfo {volume} {16}},\ \bibinfo {pages} {4121} (\bibinfo {year}
  {2013})}\BibitemShut {NoStop}%
\bibitem [{\citenamefont {Goduljan}\ \emph {et~al.}(2014)\citenamefont
  {Goduljan}, \citenamefont {Juarez}, \citenamefont {Mohammadzadeh},
  \citenamefont {Quaino}, \citenamefont {Santos},\ and\ \citenamefont
  {Schmickler}}]{schmickler:ec:14}%
  \BibitemOpen
  \bibfield  {author} {\bibinfo {author} {\bibfnamefont {Aleksej}\ \bibnamefont
  {Goduljan}}, \bibinfo {author} {\bibfnamefont {Fernanda}\ \bibnamefont
  {Juarez}}, \bibinfo {author} {\bibfnamefont {Leila}\ \bibnamefont
  {Mohammadzadeh}}, \bibinfo {author} {\bibfnamefont {Paola}\ \bibnamefont
  {Quaino}}, \bibinfo {author} {\bibfnamefont {Elizabeth}\ \bibnamefont
  {Santos}}, \ and\ \bibinfo {author} {\bibfnamefont {Wolfgang}\ \bibnamefont
  {Schmickler}},\ }\bibfield  {title} {\enquote {\bibinfo {title} {Screening of
  ions in carbon and gold nanotubes {--} a theoretical study},}\ }\href
  {\doibase 10.1016/j.elecom.2014.05.014} {\bibfield  {journal} {\bibinfo
  {journal} {Electrochem. Comm.}\ }\textbf {\bibinfo {volume} {45}},\ \bibinfo
  {pages} {48--51} (\bibinfo {year} {2014})}\BibitemShut {NoStop}%
\bibitem [{\citenamefont {Mohammadzadeh}\ \emph {et~al.}(2015)\citenamefont
  {Mohammadzadeh}, \citenamefont {Goduljan}, \citenamefont {Juarez},
  \citenamefont {Quaino}, \citenamefont {Santos},\ and\ \citenamefont
  {Schmickler}}]{schmickler:ea:15}%
  \BibitemOpen
  \bibfield  {author} {\bibinfo {author} {\bibfnamefont {Leila}\ \bibnamefont
  {Mohammadzadeh}}, \bibinfo {author} {\bibfnamefont {Aleksej}\ \bibnamefont
  {Goduljan}}, \bibinfo {author} {\bibfnamefont {Fernanda}\ \bibnamefont
  {Juarez}}, \bibinfo {author} {\bibfnamefont {Paola}\ \bibnamefont {Quaino}},
  \bibinfo {author} {\bibfnamefont {Elizabeth}\ \bibnamefont {Santos}}, \ and\
  \bibinfo {author} {\bibfnamefont {Wolfgang}\ \bibnamefont {Schmickler}},\
  }\bibfield  {title} {\enquote {\bibinfo {title} {Nanotubes for charge storage
  {--} towards an atomistic model},}\ }\href {\doibase
  doi:10.1016/j.electacta.2014.12.031} {\bibfield  {journal} {\bibinfo
  {journal} {Electrochim. Acta}\ }\textbf {\bibinfo {volume} {162}},\ \bibinfo
  {pages} {11--16} (\bibinfo {year} {2015})}\BibitemShut {NoStop}%
\bibitem [{\citenamefont {Lee}\ \emph {et~al.}(2016)\citenamefont {Lee},
  \citenamefont {Vella},\ and\ \citenamefont {Goriely}}]{vella2016quantum}%
  \BibitemOpen
  \bibfield  {author} {\bibinfo {author} {\bibfnamefont {Alpha~A.}\
  \bibnamefont {Lee}}, \bibinfo {author} {\bibfnamefont {Dominic}\ \bibnamefont
  {Vella}}, \ and\ \bibinfo {author} {\bibfnamefont {Alain}\ \bibnamefont
  {Goriely}},\ }\bibfield  {title} {\enquote {\bibinfo {title} {Quantum
  capacitance modifies interionic interactions in semiconducting nanopores},}\
  }\href@noop {} {\bibfield  {journal} {\bibinfo  {journal} {EPL (Europhysics
  Letters)}\ }\textbf {\bibinfo {volume} {113}},\ \bibinfo {pages} {38005}
  (\bibinfo {year} {2016})}\BibitemShut {NoStop}%
\bibitem [{\citenamefont {Merlet}\ \emph {et~al.}(2012)\citenamefont {Merlet},
  \citenamefont {Rotenberg}, \citenamefont {Madden}, \citenamefont {Taberna},
  \citenamefont {Simon}, \citenamefont {Gogotsi},\ and\ \citenamefont
  {Salanne}}]{merlet2012molecular}%
  \BibitemOpen
  \bibfield  {author} {\bibinfo {author} {\bibfnamefont {C{\'{e}}line}\
  \bibnamefont {Merlet}}, \bibinfo {author} {\bibfnamefont {Benjamin}\
  \bibnamefont {Rotenberg}}, \bibinfo {author} {\bibfnamefont {Paul~A.}\
  \bibnamefont {Madden}}, \bibinfo {author} {\bibfnamefont {Pierre-Louis}\
  \bibnamefont {Taberna}}, \bibinfo {author} {\bibfnamefont {Patrice}\
  \bibnamefont {Simon}}, \bibinfo {author} {\bibfnamefont {Yury}\ \bibnamefont
  {Gogotsi}}, \ and\ \bibinfo {author} {\bibfnamefont {Mathieu}\ \bibnamefont
  {Salanne}},\ }\bibfield  {title} {\enquote {\bibinfo {title} {On the
  molecular origin of supercapacitance in nanoporous carbon electrodes},}\
  }\href@noop {} {\bibfield  {journal} {\bibinfo  {journal} {Nature Mater.}\
  }\textbf {\bibinfo {volume} {11}},\ \bibinfo {pages} {306--310} (\bibinfo
  {year} {2012})}\BibitemShut {NoStop}%
\bibitem [{\citenamefont {Merlet}\ \emph {et~al.}(2013)\citenamefont {Merlet},
  \citenamefont {P{\'{e}}an}, \citenamefont {Rotenberg}, \citenamefont
  {Madden}, \citenamefont {Daffos}, \citenamefont {Taberna}, \citenamefont
  {Simon},\ and\ \citenamefont {Salanne}}]{merlet2013highly}%
  \BibitemOpen
  \bibfield  {author} {\bibinfo {author} {\bibfnamefont {C{\'{e}}line}\
  \bibnamefont {Merlet}}, \bibinfo {author} {\bibfnamefont {Clarisse}\
  \bibnamefont {P{\'{e}}an}}, \bibinfo {author} {\bibfnamefont {Benjamin}\
  \bibnamefont {Rotenberg}}, \bibinfo {author} {\bibfnamefont {Paul~A.}\
  \bibnamefont {Madden}}, \bibinfo {author} {\bibfnamefont {Barbara}\
  \bibnamefont {Daffos}}, \bibinfo {author} {\bibfnamefont {P.~L.}\
  \bibnamefont {Taberna}}, \bibinfo {author} {\bibfnamefont {Patrice}\
  \bibnamefont {Simon}}, \ and\ \bibinfo {author} {\bibfnamefont {Mathieu}\
  \bibnamefont {Salanne}},\ }\bibfield  {title} {\enquote {\bibinfo {title}
  {Highly confined ions store charge more efficiently in supercapacitors},}\
  }\href@noop {} {\bibfield  {journal} {\bibinfo  {journal} {Nature Comm.}\
  }\textbf {\bibinfo {volume} {4}},\ \bibinfo {pages} {2701} (\bibinfo {year}
  {2013})}\BibitemShut {NoStop}%
\bibitem [{\citenamefont {Wu}\ \emph {et~al.}(2011)\citenamefont {Wu},
  \citenamefont {Huang}, \citenamefont {Meunier}, \citenamefont {Sumpter},\
  and\ \citenamefont {Qiao}}]{wu2011complex}%
  \BibitemOpen
  \bibfield  {author} {\bibinfo {author} {\bibfnamefont {P.}~\bibnamefont
  {Wu}}, \bibinfo {author} {\bibfnamefont {J.}~\bibnamefont {Huang}}, \bibinfo
  {author} {\bibfnamefont {V.}~\bibnamefont {Meunier}}, \bibinfo {author}
  {\bibfnamefont {B.~G.}\ \bibnamefont {Sumpter}}, \ and\ \bibinfo {author}
  {\bibfnamefont {R.}~\bibnamefont {Qiao}},\ }\bibfield  {title} {\enquote
  {\bibinfo {title} {Complex capacitance scaling in ionic liquids-filled
  nanopores},}\ }\href@noop {} {\bibfield  {journal} {\bibinfo  {journal} {ACS
  Nano}\ }\textbf {\bibinfo {volume} {5}},\ \bibinfo {pages} {9044} (\bibinfo
  {year} {2011})}\BibitemShut {NoStop}%
\bibitem [{\citenamefont {Qiao}\ \emph {et~al.}(2012)\citenamefont {Qiao},
  \citenamefont {Meunier},\ and\ \citenamefont {Sumpter}}]{qiao2012voltage}%
  \BibitemOpen
  \bibfield  {author} {\bibinfo {author} {\bibfnamefont {R.}~\bibnamefont
  {Qiao}}, \bibinfo {author} {\bibfnamefont {V.}~\bibnamefont {Meunier}}, \
  and\ \bibinfo {author} {\bibfnamefont {B.~G.}\ \bibnamefont {Sumpter}},\
  }\bibfield  {title} {\enquote {\bibinfo {title} {Voltage dependent charge
  storage modes and capacity in subnanometer pores},}\ }\href@noop {}
  {\bibfield  {journal} {\bibinfo  {journal} {J. Phys. Chem. Lett.}\ }\textbf
  {\bibinfo {volume} {3}},\ \bibinfo {pages} {1732--1737} (\bibinfo {year}
  {2012})}\BibitemShut {NoStop}%
\bibitem [{\citenamefont {Vatamanu}\ \emph {et~al.}(2013)\citenamefont
  {Vatamanu}, \citenamefont {Hu}, \citenamefont {Bedrov}, \citenamefont
  {Perez},\ and\ \citenamefont {Gogotsi}}]{vatamanu2013increasing}%
  \BibitemOpen
  \bibfield  {author} {\bibinfo {author} {\bibfnamefont {Jenel}\ \bibnamefont
  {Vatamanu}}, \bibinfo {author} {\bibfnamefont {Zongzhi}\ \bibnamefont {Hu}},
  \bibinfo {author} {\bibfnamefont {Dmitry}\ \bibnamefont {Bedrov}}, \bibinfo
  {author} {\bibfnamefont {Carlos}\ \bibnamefont {Perez}}, \ and\ \bibinfo
  {author} {\bibfnamefont {Yury}\ \bibnamefont {Gogotsi}},\ }\bibfield  {title}
  {\enquote {\bibinfo {title} {Increasing energy storage in electrochemical
  capacitors with ionic liquid electrolytes and nanostructured carbon
  electrodes},}\ }\href@noop {} {\bibfield  {journal} {\bibinfo  {journal} {J.
  Phys. Chem. Lett}\ }\textbf {\bibinfo {volume} {4}},\ \bibinfo {pages}
  {2829--2837} (\bibinfo {year} {2013})}\BibitemShut {NoStop}%
\bibitem [{\citenamefont {Mysyk}\ \emph {et~al.}(2009)\citenamefont {Mysyk},
  \citenamefont {Raymundo-Pi{\~{n}}ero},\ and\ \citenamefont
  {B{\'{e}}guin}}]{mysyk:ec:09}%
  \BibitemOpen
  \bibfield  {author} {\bibinfo {author} {\bibfnamefont {R.}~\bibnamefont
  {Mysyk}}, \bibinfo {author} {\bibfnamefont {E.}~\bibnamefont
  {Raymundo-Pi{\~{n}}ero}}, \ and\ \bibinfo {author} {\bibfnamefont
  {F.}~\bibnamefont {B{\'{e}}guin}},\ }\bibfield  {title} {\enquote {\bibinfo
  {title} {Saturation of subnanometer pores in an electric double-layer
  capacitor},}\ }\href@noop {} {\bibfield  {journal} {\bibinfo  {journal}
  {Electrochem. Comm.}\ }\textbf {\bibinfo {volume} {11}},\ \bibinfo {pages}
  {554--556} (\bibinfo {year} {2009})}\BibitemShut {NoStop}%
\bibitem [{\citenamefont {Kondrat}\ and\ \citenamefont
  {Kornyshev}(2013)}]{kondrat:jpcc:13}%
  \BibitemOpen
  \bibfield  {author} {\bibinfo {author} {\bibfnamefont {S.}~\bibnamefont
  {Kondrat}}\ and\ \bibinfo {author} {\bibfnamefont {A.}~\bibnamefont
  {Kornyshev}},\ }\bibfield  {title} {\enquote {\bibinfo {title} {Charging
  dynamics and optimization of nano-porous supercapacitors},}\ }\href@noop {}
  {\bibfield  {journal} {\bibinfo  {journal} {J. Phys. Chem. C}\ }\textbf
  {\bibinfo {volume} {117}},\ \bibinfo {pages} {12399--12406} (\bibinfo {year}
  {2013})}\BibitemShut {NoStop}%
\bibitem [{\citenamefont {Lee}\ \emph {et~al.}(2014{\natexlab{a}})\citenamefont
  {Lee}, \citenamefont {Kondrat}, \citenamefont {Oshanin},\ and\ \citenamefont
  {Kornyshev}}]{lee:nanotech:14}%
  \BibitemOpen
  \bibfield  {author} {\bibinfo {author} {\bibfnamefont {Alpha~A.}\
  \bibnamefont {Lee}}, \bibinfo {author} {\bibfnamefont {Svyatoslav}\
  \bibnamefont {Kondrat}}, \bibinfo {author} {\bibfnamefont {Gleb}\
  \bibnamefont {Oshanin}}, \ and\ \bibinfo {author} {\bibfnamefont {Alexei~A.}\
  \bibnamefont {Kornyshev}},\ }\bibfield  {title} {\enquote {\bibinfo {title}
  {Charging dynamics of supercapacitors with narrow cylindrical nanopores},}\
  }\href {\doibase 10.1088/0957-4484/25/31/315401} {\bibfield  {journal}
  {\bibinfo  {journal} {Nanotechnology}\ }\textbf {\bibinfo {volume} {25}},\
  \bibinfo {pages} {315401} (\bibinfo {year} {2014}{\natexlab{a}})}\BibitemShut
  {NoStop}%
\bibitem [{\citenamefont {Kondrat}\ \emph {et~al.}(2014)\citenamefont
  {Kondrat}, \citenamefont {Wu}, \citenamefont {Qiao},\ and\ \citenamefont
  {Kornyshev}}]{kondrat:nm:14}%
  \BibitemOpen
  \bibfield  {author} {\bibinfo {author} {\bibfnamefont {Svyatoslav}\
  \bibnamefont {Kondrat}}, \bibinfo {author} {\bibfnamefont {Peng}\
  \bibnamefont {Wu}}, \bibinfo {author} {\bibfnamefont {Rui}\ \bibnamefont
  {Qiao}}, \ and\ \bibinfo {author} {\bibfnamefont {Alexei~A.}\ \bibnamefont
  {Kornyshev}},\ }\bibfield  {title} {\enquote {\bibinfo {title} {Accelerating
  charging dynamics in subnanometre pores},}\ }\href@noop {} {\bibfield
  {journal} {\bibinfo  {journal} {Nature Mater.}\ }\textbf {\bibinfo {volume}
  {13}},\ \bibinfo {pages} {387--393} (\bibinfo {year} {2014})}\BibitemShut
  {NoStop}%
\bibitem [{\citenamefont {Lockett}\ \emph {et~al.}(2008)\citenamefont
  {Lockett}, \citenamefont {Sedev}, \citenamefont {Ralston}, \citenamefont
  {Horne},\ and\ \citenamefont {Rodopoulos}}]{lockett:jpcc:hyster:08}%
  \BibitemOpen
  \bibfield  {author} {\bibinfo {author} {\bibfnamefont {Vera}\ \bibnamefont
  {Lockett}}, \bibinfo {author} {\bibfnamefont {Rossen}\ \bibnamefont {Sedev}},
  \bibinfo {author} {\bibfnamefont {John}\ \bibnamefont {Ralston}}, \bibinfo
  {author} {\bibfnamefont {Mike}\ \bibnamefont {Horne}}, \ and\ \bibinfo
  {author} {\bibfnamefont {Theo}\ \bibnamefont {Rodopoulos}},\ }\bibfield
  {title} {\enquote {\bibinfo {title} {Differential capacitance of the
  electrical double layer in imidazolium-based ionic liquids:{\thinspace}
  influence of potential, cation size, and temperature},}\ }\href {\doibase
  10.1021/jp7100732} {\bibfield  {journal} {\bibinfo  {journal} {J. Phys. Chem.
  C}\ }\textbf {\bibinfo {volume} {112}},\ \bibinfo {pages} {7486} (\bibinfo
  {year} {2008})}\BibitemShut {NoStop}%
\bibitem [{\citenamefont {Zhou}\ \emph {et~al.}(2010)\citenamefont {Zhou},
  \citenamefont {Inoue}, \citenamefont {Iwahashi}, \citenamefont {Kanai},
  \citenamefont {Seki}, \citenamefont {Miyamae}, \citenamefont {Kim},
  \citenamefont {Katayama},\ and\ \citenamefont
  {Ouchi}}]{zhou:ec:hysterflat:10}%
  \BibitemOpen
  \bibfield  {author} {\bibinfo {author} {\bibfnamefont {Wei}\ \bibnamefont
  {Zhou}}, \bibinfo {author} {\bibfnamefont {Soya}\ \bibnamefont {Inoue}},
  \bibinfo {author} {\bibfnamefont {Takashi}\ \bibnamefont {Iwahashi}},
  \bibinfo {author} {\bibfnamefont {Kaname}\ \bibnamefont {Kanai}}, \bibinfo
  {author} {\bibfnamefont {Kazuhiko}\ \bibnamefont {Seki}}, \bibinfo {author}
  {\bibfnamefont {Takayuki}\ \bibnamefont {Miyamae}}, \bibinfo {author}
  {\bibfnamefont {Doseok}\ \bibnamefont {Kim}}, \bibinfo {author}
  {\bibfnamefont {Yasushi}\ \bibnamefont {Katayama}}, \ and\ \bibinfo {author}
  {\bibfnamefont {Yukio}\ \bibnamefont {Ouchi}},\ }\bibfield  {title} {\enquote
  {\bibinfo {title} {Double layer structure and adsorption/desorption
  hysteresis of neat ionic liquid on pt electrode surface {--}- an in-situ
  ir-visible sum-frequency generation spectroscopic study},}\ }\href {\doibase
  doi:10.1016/j.elecom.2010.03.003} {\bibfield  {journal} {\bibinfo  {journal}
  {Electrochem. Comm.}\ }\textbf {\bibinfo {volume} {12}},\ \bibinfo {pages}
  {672} (\bibinfo {year} {2010})}\BibitemShut {NoStop}%
\bibitem [{\citenamefont {Uysal}\ \emph {et~al.}(2013)\citenamefont {Uysal},
  \citenamefont {Zhou}, \citenamefont {Feng}, \citenamefont {Lee},
  \citenamefont {Li}, \citenamefont {Fenter}, \citenamefont {Cummings},
  \citenamefont {Fulvio}, \citenamefont {Dai}, \citenamefont {McDonough},\ and\
  \citenamefont {Gogotsi}}]{uysala:jpcc:hyster:13}%
  \BibitemOpen
  \bibfield  {author} {\bibinfo {author} {\bibfnamefont {Ahmet}\ \bibnamefont
  {Uysal}}, \bibinfo {author} {\bibfnamefont {Hua}\ \bibnamefont {Zhou}},
  \bibinfo {author} {\bibfnamefont {Guang}\ \bibnamefont {Feng}}, \bibinfo
  {author} {\bibfnamefont {Sang~Soo}\ \bibnamefont {Lee}}, \bibinfo {author}
  {\bibfnamefont {Song}\ \bibnamefont {Li}}, \bibinfo {author} {\bibfnamefont
  {Paul}\ \bibnamefont {Fenter}}, \bibinfo {author} {\bibfnamefont {Peter~T.}\
  \bibnamefont {Cummings}}, \bibinfo {author} {\bibfnamefont {Pasquale~F.}\
  \bibnamefont {Fulvio}}, \bibinfo {author} {\bibfnamefont {Sheng}\
  \bibnamefont {Dai}}, \bibinfo {author} {\bibfnamefont {John~K.}\ \bibnamefont
  {McDonough}}, \ and\ \bibinfo {author} {\bibfnamefont {Yury}\ \bibnamefont
  {Gogotsi}},\ }\bibfield  {title} {\enquote {\bibinfo {title} {Structural
  origins of potential dependent hysteresis at the electrified graphene/ionic
  liquid interface},}\ }\href {\doibase 10.1021/jp4111025} {\bibfield
  {journal} {\bibinfo  {journal} {J. Phys. Chem. C}\ }\textbf {\bibinfo
  {volume} {118}},\ \bibinfo {pages} {569} (\bibinfo {year}
  {2013})}\BibitemShut {NoStop}%
\bibitem [{\citenamefont {Rotenberg}\ and\ \citenamefont
  {Salanne}(2015)}]{rotenberg2015structural}%
  \BibitemOpen
  \bibfield  {author} {\bibinfo {author} {\bibfnamefont {Benjamin}\
  \bibnamefont {Rotenberg}}\ and\ \bibinfo {author} {\bibfnamefont {Mathieu}\
  \bibnamefont {Salanne}},\ }\bibfield  {title} {\enquote {\bibinfo {title}
  {Structural transitions at ionic liquid interfaces},}\ }\href@noop {}
  {\bibfield  {journal} {\bibinfo  {journal} {The Journal of Physical Chemistry
  Letters}\ }\textbf {\bibinfo {volume} {6}},\ \bibinfo {pages} {4978--4985}
  (\bibinfo {year} {2015})}\BibitemShut {NoStop}%
\bibitem [{\citenamefont {Dr{\"{u}}schler}\ \emph {et~al.}(2010)\citenamefont
  {Dr{\"{u}}schler}, \citenamefont {Huber}, \citenamefont {Passerini},\ and\
  \citenamefont {Roling}}]{druechler:jpcc:hyster:10}%
  \BibitemOpen
  \bibfield  {author} {\bibinfo {author} {\bibfnamefont {Marcel}\ \bibnamefont
  {Dr{\"{u}}schler}}, \bibinfo {author} {\bibfnamefont {Benedikt}\ \bibnamefont
  {Huber}}, \bibinfo {author} {\bibfnamefont {Stefano}\ \bibnamefont
  {Passerini}}, \ and\ \bibinfo {author} {\bibfnamefont {Bernhard}\
  \bibnamefont {Roling}},\ }\bibfield  {title} {\enquote {\bibinfo {title}
  {Hysteresis effects in the potential-dependent double layer capacitance of
  room temperature ionic liquids at a polycrystalline platinum interface},}\
  }\href {\doibase 10.1021/jp911513k} {\bibfield  {journal} {\bibinfo
  {journal} {J. Phys. Chem. C}\ }\textbf {\bibinfo {volume} {114}},\ \bibinfo
  {pages} {3614} (\bibinfo {year} {2010})}\BibitemShut {NoStop}%
\bibitem [{\citenamefont {Kiyohara}\ \emph {et~al.}(2011)\citenamefont
  {Kiyohara}, \citenamefont {Sugino},\ and\ \citenamefont
  {Asaka}}]{kiyohara2011phase}%
  \BibitemOpen
  \bibfield  {author} {\bibinfo {author} {\bibfnamefont {K.}~\bibnamefont
  {Kiyohara}}, \bibinfo {author} {\bibfnamefont {T.}~\bibnamefont {Sugino}}, \
  and\ \bibinfo {author} {\bibfnamefont {K.}~\bibnamefont {Asaka}},\ }\bibfield
   {title} {\enquote {\bibinfo {title} {Phase transition in porous
  electrodes},}\ }\href@noop {} {\bibfield  {journal} {\bibinfo  {journal} {J.
  Chem. Phys.}\ }\textbf {\bibinfo {volume} {134}},\ \bibinfo {pages} {154710}
  (\bibinfo {year} {2011})}\BibitemShut {NoStop}%
\bibitem [{\citenamefont {Xing}\ \emph {et~al.}(2013)\citenamefont {Xing},
  \citenamefont {Vatamanu}, \citenamefont {Borodin},\ and\ \citenamefont
  {Bedrov}}]{xing:jpcl:trans:13}%
  \BibitemOpen
  \bibfield  {author} {\bibinfo {author} {\bibfnamefont {Lidan}\ \bibnamefont
  {Xing}}, \bibinfo {author} {\bibfnamefont {Jenel}\ \bibnamefont {Vatamanu}},
  \bibinfo {author} {\bibfnamefont {Oleg}\ \bibnamefont {Borodin}}, \ and\
  \bibinfo {author} {\bibfnamefont {Dmitry}\ \bibnamefont {Bedrov}},\
  }\bibfield  {title} {\enquote {\bibinfo {title} {On the atomistic nature of
  capacitance enhancement generated by ionic liquid electrolyte confined in
  subnanometer pores},}\ }\href {\doibase 10.1021/jz301782f} {\bibfield
  {journal} {\bibinfo  {journal} {J. Phys. Chem. Lett.}\ }\textbf {\bibinfo
  {volume} {4}},\ \bibinfo {pages} {132} (\bibinfo {year} {2013})}\BibitemShut
  {NoStop}%
\bibitem [{\citenamefont {Vatamanu}\ \emph {et~al.}(2015)\citenamefont
  {Vatamanu}, \citenamefont {Vatamanu},\ and\ \citenamefont
  {Bedrov}}]{vatamanu:acsnano:15}%
  \BibitemOpen
  \bibfield  {author} {\bibinfo {author} {\bibfnamefont {Jenel}\ \bibnamefont
  {Vatamanu}}, \bibinfo {author} {\bibfnamefont {Mihaela}\ \bibnamefont
  {Vatamanu}}, \ and\ \bibinfo {author} {\bibfnamefont {Dmitry}\ \bibnamefont
  {Bedrov}},\ }\bibfield  {title} {\enquote {\bibinfo {title} {Non-faradic
  energy storage by room temperature ionic liquids in nanoporous electrodes},}\
  }\href {\doibase 10.1021/acsnano.5b00945} {\bibfield  {journal} {\bibinfo
  {journal} {ACS nano}\ }\textbf {\bibinfo {volume} {9}},\ \bibinfo {pages}
  {5999} (\bibinfo {year} {2015})}\BibitemShut {NoStop}%
\bibitem [{\citenamefont {Kondrat}\ \emph {et~al.}(2011)\citenamefont
  {Kondrat}, \citenamefont {Georgi}, \citenamefont {Fedorov},\ and\
  \citenamefont {Kornyshev}}]{kondrat2011superionicMC}%
  \BibitemOpen
  \bibfield  {author} {\bibinfo {author} {\bibfnamefont {S.}~\bibnamefont
  {Kondrat}}, \bibinfo {author} {\bibfnamefont {N.}~\bibnamefont {Georgi}},
  \bibinfo {author} {\bibfnamefont {M.~V.}\ \bibnamefont {Fedorov}}, \ and\
  \bibinfo {author} {\bibfnamefont {A.~A.}\ \bibnamefont {Kornyshev}},\
  }\bibfield  {title} {\enquote {\bibinfo {title} {A superionic state in
  nano-porous double-layer capacitors: insights from monte carlo
  simulations},}\ }\href@noop {} {\bibfield  {journal} {\bibinfo  {journal}
  {Phys. Chem. Chem. Phys.}\ }\textbf {\bibinfo {volume} {13}},\ \bibinfo
  {pages} {11359--11366} (\bibinfo {year} {2011})}\BibitemShut {NoStop}%
\bibitem [{\citenamefont {Kiyohara}\ \emph {et~al.}(2012)\citenamefont
  {Kiyohara}, \citenamefont {Shioyama}, \citenamefont {Sugino},\ and\
  \citenamefont {Asaka}}]{kiyohara2012phase}%
  \BibitemOpen
  \bibfield  {author} {\bibinfo {author} {\bibfnamefont {K.}~\bibnamefont
  {Kiyohara}}, \bibinfo {author} {\bibfnamefont {H.}~\bibnamefont {Shioyama}},
  \bibinfo {author} {\bibfnamefont {T.}~\bibnamefont {Sugino}}, \ and\ \bibinfo
  {author} {\bibfnamefont {K.}~\bibnamefont {Asaka}},\ }\bibfield  {title}
  {\enquote {\bibinfo {title} {Phase transition in porous electrodes. ii.
  effect of asymmetry in the ion size},}\ }\href@noop {} {\bibfield  {journal}
  {\bibinfo  {journal} {J. Chem. Phys.}\ }\textbf {\bibinfo {volume} {136}},\
  \bibinfo {pages} {094701} (\bibinfo {year} {2012})}\BibitemShut {NoStop}%
\bibitem [{\citenamefont {Kiyohara}\ \emph {et~al.}(2013)\citenamefont
  {Kiyohara}, \citenamefont {Shioyama}, \citenamefont {Sugino}, \citenamefont
  {Asaka}, \citenamefont {Soneda}, \citenamefont {Imoto},\ and\ \citenamefont
  {Kodama}}]{kiyohara2013phase}%
  \BibitemOpen
  \bibfield  {author} {\bibinfo {author} {\bibfnamefont {Kenji}\ \bibnamefont
  {Kiyohara}}, \bibinfo {author} {\bibfnamefont {Hiroshi}\ \bibnamefont
  {Shioyama}}, \bibinfo {author} {\bibfnamefont {Takushi}\ \bibnamefont
  {Sugino}}, \bibinfo {author} {\bibfnamefont {Kinji}\ \bibnamefont {Asaka}},
  \bibinfo {author} {\bibfnamefont {Yasushi}\ \bibnamefont {Soneda}}, \bibinfo
  {author} {\bibfnamefont {Kiyoaki}\ \bibnamefont {Imoto}}, \ and\ \bibinfo
  {author} {\bibfnamefont {Masaya}\ \bibnamefont {Kodama}},\ }\bibfield
  {title} {\enquote {\bibinfo {title} {Phase transition in porous electrodes.
  iii. for the case of a two component electrolyte},}\ }\href@noop {}
  {\bibfield  {journal} {\bibinfo  {journal} {J. Chem. Phys.}\ }\textbf
  {\bibinfo {volume} {138}},\ \bibinfo {pages} {234704} (\bibinfo {year}
  {2013})}\BibitemShut {NoStop}%
\bibitem [{\citenamefont {Jiang}\ \emph {et~al.}(2011)\citenamefont {Jiang},
  \citenamefont {Jin},\ and\ \citenamefont {Wu}}]{jiang2011oscillation}%
  \BibitemOpen
  \bibfield  {author} {\bibinfo {author} {\bibfnamefont {D.}~\bibnamefont
  {Jiang}}, \bibinfo {author} {\bibfnamefont {Z.}~\bibnamefont {Jin}}, \ and\
  \bibinfo {author} {\bibfnamefont {J.}~\bibnamefont {Wu}},\ }\bibfield
  {title} {\enquote {\bibinfo {title} {Oscillation of capacitance inside
  nanopores},}\ }\href@noop {} {\bibfield  {journal} {\bibinfo  {journal} {Nano
  Lett.}\ }\textbf {\bibinfo {volume} {11}},\ \bibinfo {pages} {5373} (\bibinfo
  {year} {2011})}\BibitemShut {NoStop}%
\bibitem [{\citenamefont {Jiang}\ and\ \citenamefont
  {Wu}(2014)}]{jiang:nanoscale:14}%
  \BibitemOpen
  \bibfield  {author} {\bibinfo {author} {\bibfnamefont {De-en}\ \bibnamefont
  {Jiang}}\ and\ \bibinfo {author} {\bibfnamefont {Jianzhong}\ \bibnamefont
  {Wu}},\ }\bibfield  {title} {\enquote {\bibinfo {title} {Unusual effects of
  solvent polarity on capacitance for organic electrolytes in a nanoporous
  electrode},}\ }\href {\doibase 10.1039/c4nr00046c} {\bibfield  {journal}
  {\bibinfo  {journal} {Nanoscale}\ }\textbf {\bibinfo {volume} {6}},\ \bibinfo
  {pages} {5545} (\bibinfo {year} {2014})}\BibitemShut {NoStop}%
\bibitem [{\citenamefont {Jover}\ \emph {et~al.}(2014)\citenamefont {Jover},
  \citenamefont {Lugo}, \citenamefont {Toulhoat}, \citenamefont {Simon},\ and\
  \citenamefont {De~Bruin}}]{jover2014screening}%
  \BibitemOpen
  \bibfield  {author} {\bibinfo {author} {\bibfnamefont {Julio~F.}\
  \bibnamefont {Jover}}, \bibinfo {author} {\bibfnamefont {Rafael}\
  \bibnamefont {Lugo}}, \bibinfo {author} {\bibfnamefont {Herv{\'{e}}}\
  \bibnamefont {Toulhoat}}, \bibinfo {author} {\bibfnamefont {Patrice}\
  \bibnamefont {Simon}}, \ and\ \bibinfo {author} {\bibfnamefont {Theodorus}\
  \bibnamefont {De~Bruin}},\ }\bibfield  {title} {\enquote {\bibinfo {title}
  {Screening methodology for the efficient pairing of ionic liquids and
  carbonaceous electrodes applied to electric energy storage},}\ }\href@noop {}
  {\bibfield  {journal} {\bibinfo  {journal} {J. Phys. Chem. C}\ }\textbf
  {\bibinfo {volume} {118}},\ \bibinfo {pages} {864--872} (\bibinfo {year}
  {2014})}\BibitemShut {NoStop}%
\bibitem [{Note1()}]{Note1}%
  \BibitemOpen
  \bibinfo {note} {From ref.~\cite {jover2014screening}, the energy to transfer
  an ion pair from the bulk to vacuum $\approx 50-85 k_B T$, and the energy to
  dissociate an ion pair at vacuum $\approx 80-140 k_B T$, therefore the
  desolvation energy of a single ion $\approx 65 -115k_BT$.}\BibitemShut
  {Stop}%
\bibitem [{Note2()}]{Note2}%
  \BibitemOpen
  \bibinfo {note} {Estimate obtained from Ref.~\cite {jover2014screening},
  where a cylindrical geometry is studied. Our estimate is obtained by taking
  the large radius asymptotic value of the ion-wall van der Waals interaction
  energy, \protect \emph {i.e.} $E_{\protect \mathrm {vdw,slit}} \approx 2
  E_{\protect \mathrm {vdw,cylinder}} (R \rightarrow \infty )$, as the $R
  \rightarrow \infty $ asymptotic value of the van der Waals interaction energy
  corresponds to the interaction energy between an ion and one side of the
  pore.}\BibitemShut {Stop}%
\bibitem [{\citenamefont {Griffin}\ \emph {et~al.}(2015)\citenamefont
  {Griffin}, \citenamefont {Forse}, \citenamefont {Tsai}, \citenamefont
  {Taberna}, \citenamefont {Simon},\ and\ \citenamefont
  {Grey}}]{griffin:natmat:15}%
  \BibitemOpen
  \bibfield  {author} {\bibinfo {author} {\bibfnamefont {John~M.}\ \bibnamefont
  {Griffin}}, \bibinfo {author} {\bibfnamefont {Alexander~C.}\ \bibnamefont
  {Forse}}, \bibinfo {author} {\bibfnamefont {Wan-Yu}\ \bibnamefont {Tsai}},
  \bibinfo {author} {\bibfnamefont {Pierre-Louis}\ \bibnamefont {Taberna}},
  \bibinfo {author} {\bibfnamefont {Patrice}\ \bibnamefont {Simon}}, \ and\
  \bibinfo {author} {\bibfnamefont {Clare~P.}\ \bibnamefont {Grey}},\
  }\bibfield  {title} {\enquote {\bibinfo {title} {In situ nmr and
  electrochemical quartz crystal microbalance techniques reveal the structure
  of the electrical double layer in supercapacitors},}\ }\href {\doibase
  10.1038/nmat4318} {\bibfield  {journal} {\bibinfo  {journal} {Nat. Mater.}\
  }\textbf {\bibinfo {volume} {14}},\ \bibinfo {pages} {812} (\bibinfo {year}
  {2015})}\BibitemShut {NoStop}%
\bibitem [{\citenamefont {Forse}\ \emph {et~al.}(2015)\citenamefont {Forse},
  \citenamefont {Griffin}, \citenamefont {Merlet}, \citenamefont {Bayley},
  \citenamefont {Wang}, \citenamefont {Simon},\ and\ \citenamefont
  {Grey}}]{force:jacs:NMRDyn}%
  \BibitemOpen
  \bibfield  {author} {\bibinfo {author} {\bibfnamefont {Alexander~C.}\
  \bibnamefont {Forse}}, \bibinfo {author} {\bibfnamefont {John~M.}\
  \bibnamefont {Griffin}}, \bibinfo {author} {\bibfnamefont {C{\'{e}}line}\
  \bibnamefont {Merlet}}, \bibinfo {author} {\bibfnamefont {Paul~M.}\
  \bibnamefont {Bayley}}, \bibinfo {author} {\bibfnamefont {Hao}\ \bibnamefont
  {Wang}}, \bibinfo {author} {\bibfnamefont {Patrice}\ \bibnamefont {Simon}}, \
  and\ \bibinfo {author} {\bibfnamefont {Clare~P.}\ \bibnamefont {Grey}},\
  }\bibfield  {title} {\enquote {\bibinfo {title} {Nmr study of ion dynamics
  and charge storage in ionic liquid supercapacitors},}\ }\href {\doibase
  10.1021/jacs.5b03958} {\bibfield  {journal} {\bibinfo  {journal} {J. Am.
  Chem. Soc.}\ }\textbf {\bibinfo {volume} {137}},\ \bibinfo {pages}
  {7231--7242} (\bibinfo {year} {2015})}\BibitemShut {NoStop}%
\bibitem [{\citenamefont {Rochester}\ \emph {et~al.}(2016)\citenamefont
  {Rochester}, \citenamefont {Kondrat}, \citenamefont {Pruessner},\ and\
  \citenamefont {Kornyshev}}]{rochester:1d}%
  \BibitemOpen
  \bibfield  {author} {\bibinfo {author} {\bibfnamefont {C.~C.}\ \bibnamefont
  {Rochester}}, \bibinfo {author} {\bibfnamefont {S.}~\bibnamefont {Kondrat}},
  \bibinfo {author} {\bibfnamefont {G.}~\bibnamefont {Pruessner}}, \ and\
  \bibinfo {author} {\bibfnamefont {A.~A.}\ \bibnamefont {Kornyshev}},\
  }\bibfield  {title} {\enquote {\bibinfo {title} {Charging ultra-nanoporous
  electrodes with size-asymmetric ions assisted by apolar solvent},}\ }\href
  {\doibase 10.1021/acs.jpcc.5b12730} {\bibfield  {journal} {\bibinfo
  {journal} {J. Phys. Chem. C}\ } (\bibinfo {year} {2016}),\
  10.1021/acs.jpcc.5b12730},\ \bibinfo {note} {in press}\BibitemShut {NoStop}%
\bibitem [{\citenamefont {Lee}\ \emph {et~al.}(2014{\natexlab{b}})\citenamefont
  {Lee}, \citenamefont {Kondrat},\ and\ \citenamefont
  {Kornyshev}}]{leePRL2014}%
  \BibitemOpen
  \bibfield  {author} {\bibinfo {author} {\bibfnamefont {Alpha~A.}\
  \bibnamefont {Lee}}, \bibinfo {author} {\bibfnamefont {S.}~\bibnamefont
  {Kondrat}}, \ and\ \bibinfo {author} {\bibfnamefont {Alexei~A.}\ \bibnamefont
  {Kornyshev}},\ }\bibfield  {title} {\enquote {\bibinfo {title} {Single-file
  charge storage in conducting nanopores},}\ }\href@noop {} {\bibfield
  {journal} {\bibinfo  {journal} {Phys. Rev. Lett}\ }\textbf {\bibinfo {volume}
  {113}},\ \bibinfo {pages} {48701} (\bibinfo {year}
  {2014}{\natexlab{b}})}\BibitemShut {NoStop}%
\bibitem [{\citenamefont {Kondrat}\ and\ \citenamefont
  {Kornyshev}(2016)}]{kondrat:nh:16}%
  \BibitemOpen
  \bibfield  {author} {\bibinfo {author} {\bibfnamefont {Svyatoslav}\
  \bibnamefont {Kondrat}}\ and\ \bibinfo {author} {\bibfnamefont {Alexei}\
  \bibnamefont {Kornyshev}},\ }\bibfield  {title} {\enquote {\bibinfo {title}
  {Pressing a spring: What does it take to maximize the energy storage in
  nanoporous supercapacitors?}}\ }\href {\doibase 10.1039/C5NH00004A}
  {\bibfield  {journal} {\bibinfo  {journal} {Nanoscale Horiz.}\ }\textbf
  {\bibinfo {volume} {1}},\ \bibinfo {pages} {45--52} (\bibinfo {year}
  {2016})}\BibitemShut {NoStop}%
\bibitem [{\citenamefont {Feng}\ and\ \citenamefont
  {Cummings}(2011)}]{feng2011supercapacitor}%
  \BibitemOpen
  \bibfield  {author} {\bibinfo {author} {\bibfnamefont {G.}~\bibnamefont
  {Feng}}\ and\ \bibinfo {author} {\bibfnamefont {P.~T.}\ \bibnamefont
  {Cummings}},\ }\bibfield  {title} {\enquote {\bibinfo {title} {Supercapacitor
  capacitance exhibits oscillatory behavior as a function of nanopore size},}\
  }\href@noop {} {\bibfield  {journal} {\bibinfo  {journal} {J. Phys. Chem.
  Lett}\ }\textbf {\bibinfo {volume} {2}},\ \bibinfo {pages} {2859} (\bibinfo
  {year} {2011})}\BibitemShut {NoStop}%
\bibitem [{\citenamefont {Pizio}\ \emph {et~al.}(2012)\citenamefont {Pizio},
  \citenamefont {Sokolowski},\ and\ \citenamefont
  {Sokolowska}}]{pizio2012electric}%
  \BibitemOpen
  \bibfield  {author} {\bibinfo {author} {\bibfnamefont {O.}~\bibnamefont
  {Pizio}}, \bibinfo {author} {\bibfnamefont {S.}~\bibnamefont {Sokolowski}}, \
  and\ \bibinfo {author} {\bibfnamefont {Z.}~\bibnamefont {Sokolowska}},\
  }\bibfield  {title} {\enquote {\bibinfo {title} {Electric double layer
  capacitance of restricted primitive model for an ionic fluid in slit-like
  nanopores: A density functional approach},}\ }\href@noop {} {\bibfield
  {journal} {\bibinfo  {journal} {J. Chem. Phys.}\ }\textbf {\bibinfo {volume}
  {137}},\ \bibinfo {pages} {234705} (\bibinfo {year} {2012})}\BibitemShut
  {NoStop}%
\bibitem [{Note3()}]{Note3}%
  \BibitemOpen
  \bibinfo {note} {We note however that this reasoning is valid for pore-width
  independent (or only weakly dependent) resolvation energies. Although this is
  not generally the case, it seems reasonable to expect it to hold for small
  variations of the slit width $d < L < 2d$ considered in this
  work.}\BibitemShut {Stop}%
\bibitem [{\citenamefont {Griffin}\ \emph {et~al.}(2014)\citenamefont
  {Griffin}, \citenamefont {Forse}, \citenamefont {Wang}, \citenamefont
  {Trease}, \citenamefont {Taberna}, \citenamefont {Simon},\ and\ \citenamefont
  {Grey}}]{griffin:fd:14}%
  \BibitemOpen
  \bibfield  {author} {\bibinfo {author} {\bibfnamefont {John~M.}\ \bibnamefont
  {Griffin}}, \bibinfo {author} {\bibfnamefont {Alexander~C.}\ \bibnamefont
  {Forse}}, \bibinfo {author} {\bibfnamefont {Hao}\ \bibnamefont {Wang}},
  \bibinfo {author} {\bibfnamefont {Nicole~M.}\ \bibnamefont {Trease}},
  \bibinfo {author} {\bibfnamefont {Pierre-Louis}\ \bibnamefont {Taberna}},
  \bibinfo {author} {\bibfnamefont {Patrice}\ \bibnamefont {Simon}}, \ and\
  \bibinfo {author} {\bibfnamefont {Clare~P.}\ \bibnamefont {Grey}},\
  }\bibfield  {title} {\enquote {\bibinfo {title} {Ion counting in
  supercapacitor electrodes using nmr spectroscopy},}\ }\href {\doibase
  10.1039/C4FD00138A} {\bibfield  {journal} {\bibinfo  {journal} {Faraday
  Discuss.}\ } (\bibinfo {year} {2014}),\ 10.1039/C4FD00138A},\ \bibinfo {note}
  {accepted}\BibitemShut {NoStop}%
\bibitem [{\citenamefont {Kondrat}\ \emph {et~al.}(2012)\citenamefont
  {Kondrat}, \citenamefont {P{\'{e}}rez}, \citenamefont {Presser},
  \citenamefont {Gogotsi},\ and\ \citenamefont {Kornyshev}}]{kondrat:ees:12}%
  \BibitemOpen
  \bibfield  {author} {\bibinfo {author} {\bibfnamefont {S.}~\bibnamefont
  {Kondrat}}, \bibinfo {author} {\bibfnamefont {C.~R.}\ \bibnamefont
  {P{\'{e}}rez}}, \bibinfo {author} {\bibfnamefont {V.}~\bibnamefont
  {Presser}}, \bibinfo {author} {\bibfnamefont {Y.}~\bibnamefont {Gogotsi}}, \
  and\ \bibinfo {author} {\bibfnamefont {A.~A.}\ \bibnamefont {Kornyshev}},\
  }\bibfield  {title} {\enquote {\bibinfo {title} {Effect of pore size and its
  dispersity on the energy storage in nanoporous supercapacitors},}\
  }\href@noop {} {\bibfield  {journal} {\bibinfo  {journal} {Energy Environ.
  Sci.}\ }\textbf {\bibinfo {volume} {5}},\ \bibinfo {pages} {6474--6479}
  (\bibinfo {year} {2012})}\BibitemShut {NoStop}%
\bibitem [{\citenamefont {Wang}\ \emph {et~al.}(2013)\citenamefont {Wang},
  \citenamefont {Forse}, \citenamefont {Griffin}, \citenamefont {Trease},
  \citenamefont {Trognko}, \citenamefont {Taberna}, \citenamefont {Simon},\
  and\ \citenamefont {Grey}}]{wang:jacs:NMRsupercap:13}%
  \BibitemOpen
  \bibfield  {author} {\bibinfo {author} {\bibfnamefont {Hao}\ \bibnamefont
  {Wang}}, \bibinfo {author} {\bibfnamefont {Alexander~C.}\ \bibnamefont
  {Forse}}, \bibinfo {author} {\bibfnamefont {John~M.}\ \bibnamefont
  {Griffin}}, \bibinfo {author} {\bibfnamefont {Nicole~M.}\ \bibnamefont
  {Trease}}, \bibinfo {author} {\bibfnamefont {Lorie}\ \bibnamefont {Trognko}},
  \bibinfo {author} {\bibfnamefont {Pierre-Louis}\ \bibnamefont {Taberna}},
  \bibinfo {author} {\bibfnamefont {Patrice}\ \bibnamefont {Simon}}, \ and\
  \bibinfo {author} {\bibfnamefont {Clare~P.}\ \bibnamefont {Grey}},\
  }\bibfield  {title} {\enquote {\bibinfo {title} {In situ nmr spectroscopy of
  supercapacitors: Insight into the charge storage mechanism},}\ }\href
  {\doibase 10.1021/ja410287s} {\bibfield  {journal} {\bibinfo  {journal} {J.
  Am. Chem. Soc.}\ }\textbf {\bibinfo {volume} {135}},\ \bibinfo {pages}
  {18968} (\bibinfo {year} {2013})}\BibitemShut {NoStop}%
\bibitem [{\citenamefont {Forse}\ \emph {et~al.}(2013)\citenamefont {Forse},
  \citenamefont {Griffin}, \citenamefont {Wang}, \citenamefont {Trease},
  \citenamefont {Presser}, \citenamefont {Gogotsi}, \citenamefont {Simone},\
  and\ \citenamefont {Grey}}]{forse:pccp:13}%
  \BibitemOpen
  \bibfield  {author} {\bibinfo {author} {\bibfnamefont {Alexander~C.}\
  \bibnamefont {Forse}}, \bibinfo {author} {\bibfnamefont {John~M.}\
  \bibnamefont {Griffin}}, \bibinfo {author} {\bibfnamefont {Hao}\ \bibnamefont
  {Wang}}, \bibinfo {author} {\bibfnamefont {Nicole~M.}\ \bibnamefont
  {Trease}}, \bibinfo {author} {\bibfnamefont {Volker}\ \bibnamefont
  {Presser}}, \bibinfo {author} {\bibfnamefont {Yury}\ \bibnamefont {Gogotsi}},
  \bibinfo {author} {\bibfnamefont {Patrice}\ \bibnamefont {Simone}}, \ and\
  \bibinfo {author} {\bibfnamefont {Clare~P.}\ \bibnamefont {Grey}},\
  }\bibfield  {title} {\enquote {\bibinfo {title} {Nuclear magnetic resonance
  study of ion adsorption on microporous carbide-derived carbon},}\ }\href
  {\doibase 10.1039/C3CP51210J} {\bibfield  {journal} {\bibinfo  {journal}
  {Phys Chem Chem Phys}\ }\textbf {\bibinfo {volume} {15}},\ \bibinfo {pages}
  {7722} (\bibinfo {year} {2013})}\BibitemShut {NoStop}%
\bibitem [{\citenamefont {Levi}\ \emph {et~al.}(2009)\citenamefont {Levi},
  \citenamefont {Levy}, \citenamefont {Aurbach},\ and\ \citenamefont
  {Maier}}]{levi:nm:eqcm:09}%
  \BibitemOpen
  \bibfield  {author} {\bibinfo {author} {\bibfnamefont {Mikhael~D.}\
  \bibnamefont {Levi}}, \bibinfo {author} {\bibfnamefont {Grigory
  Salitraand~Naomi}\ \bibnamefont {Levy}}, \bibinfo {author} {\bibfnamefont
  {Doron}\ \bibnamefont {Aurbach}}, \ and\ \bibinfo {author} {\bibfnamefont
  {Joachim}\ \bibnamefont {Maier}},\ }\bibfield  {title} {\enquote {\bibinfo
  {title} {Application of a quartz-crystal microbalance to measure ionic fluxes
  in microporous carbons for energy storage},}\ }\href {\doibase
  10.1038/nmat2559} {\bibfield  {journal} {\bibinfo  {journal} {Nat. Mater.}\
  }\textbf {\bibinfo {volume} {8}},\ \bibinfo {pages} {872} (\bibinfo {year}
  {2009})}\BibitemShut {NoStop}%
\bibitem [{\citenamefont {Levi}\ \emph {et~al.}(2013)\citenamefont {Levi},
  \citenamefont {Sigalov}, \citenamefont {Aurbach},\ and\ \citenamefont
  {Daikhin}}]{levi:jpcc:13:EQCM}%
  \BibitemOpen
  \bibfield  {author} {\bibinfo {author} {\bibfnamefont {Mikhael~D.}\
  \bibnamefont {Levi}}, \bibinfo {author} {\bibfnamefont {Sergey}\ \bibnamefont
  {Sigalov}}, \bibinfo {author} {\bibfnamefont {Doron}\ \bibnamefont
  {Aurbach}}, \ and\ \bibinfo {author} {\bibfnamefont {Leonid}\ \bibnamefont
  {Daikhin}},\ }\bibfield  {title} {\enquote {\bibinfo {title} {In situ
  electrochemical quartz crystal admittance methodology for tracking
  compositional and mechanical changes in porous carbon electrodes},}\ }\href
  {\doibase 10.1021/jp403065y} {\bibfield  {journal} {\bibinfo  {journal} {J.
  Phys. Chem. C}\ }\textbf {\bibinfo {volume} {117}},\ \bibinfo {pages}
  {4876--14889} (\bibinfo {year} {2013})}\BibitemShut {NoStop}%
\bibitem [{\citenamefont {Tsai}\ \emph {et~al.}(2014)\citenamefont {Tsai},
  \citenamefont {Taberna},\ and\ \citenamefont {Simon}}]{tsai:jacs:14}%
  \BibitemOpen
  \bibfield  {author} {\bibinfo {author} {\bibfnamefont {Wan-Yu}\ \bibnamefont
  {Tsai}}, \bibinfo {author} {\bibfnamefont {Pierre-Louis}\ \bibnamefont
  {Taberna}}, \ and\ \bibinfo {author} {\bibfnamefont {Patrice}\ \bibnamefont
  {Simon}},\ }\bibfield  {title} {\enquote {\bibinfo {title} {Electrochemical
  quartz crystal microbalance (eqcm) study of ion dynamics in nanoporous
  carbons.}}\ }\href {\doibase 10.1021/ja503449w} {\bibfield  {journal}
  {\bibinfo  {journal} {J. Am. Chem. Soc.}\ }\textbf {\bibinfo {volume}
  {136}},\ \bibinfo {pages} {8722} (\bibinfo {year} {2014})}\BibitemShut
  {NoStop}%
\bibitem [{\citenamefont {Xia}\ \emph {et~al.}(2009)\citenamefont {Xia},
  \citenamefont {Chen}, \citenamefont {Li},\ and\ \citenamefont
  {Tao}}]{tao:natnano:graphene:09}%
  \BibitemOpen
  \bibfield  {author} {\bibinfo {author} {\bibfnamefont {Jilin}\ \bibnamefont
  {Xia}}, \bibinfo {author} {\bibfnamefont {Fang}\ \bibnamefont {Chen}},
  \bibinfo {author} {\bibfnamefont {Jinghong}\ \bibnamefont {Li}}, \ and\
  \bibinfo {author} {\bibfnamefont {Nongjian}\ \bibnamefont {Tao}},\ }\bibfield
   {title} {\enquote {\bibinfo {title} {Measurement of the quantum capacitance
  of graphene},}\ }\href {\doibase 10.1038/nnano.2009.177} {\bibfield
  {journal} {\bibinfo  {journal} {Nat. Nanotech.}\ }\textbf {\bibinfo {volume}
  {4}},\ \bibinfo {pages} {505} (\bibinfo {year} {2009})}\BibitemShut {NoStop}%
\bibitem [{\citenamefont {Stoller}\ \emph {et~al.}(2011)\citenamefont
  {Stoller}, \citenamefont {Magnuson}, \citenamefont {Zhu}, \citenamefont
  {Murali}, \citenamefont {Suk}, \citenamefont {Piner},\ and\ \citenamefont
  {Ruoff}}]{stoller:ees:graphene:11}%
  \BibitemOpen
  \bibfield  {author} {\bibinfo {author} {\bibfnamefont {Meryl~D.}\
  \bibnamefont {Stoller}}, \bibinfo {author} {\bibfnamefont {Carl~W.}\
  \bibnamefont {Magnuson}}, \bibinfo {author} {\bibfnamefont {Yanwu}\
  \bibnamefont {Zhu}}, \bibinfo {author} {\bibfnamefont {Shanthi}\ \bibnamefont
  {Murali}}, \bibinfo {author} {\bibfnamefont {Ji~Won}\ \bibnamefont {Suk}},
  \bibinfo {author} {\bibfnamefont {Richard}\ \bibnamefont {Piner}}, \ and\
  \bibinfo {author} {\bibfnamefont {Rodney~S.}\ \bibnamefont {Ruoff}},\
  }\bibfield  {title} {\enquote {\bibinfo {title} {Interfacial capacitance of
  single layer graphene},}\ }\href {\doibase 10.1039/C1EE02322E} {\bibfield
  {journal} {\bibinfo  {journal} {Energy Environ. Sci.}\ }\textbf {\bibinfo
  {volume} {4}},\ \bibinfo {pages} {4685} (\bibinfo {year} {2011})}\BibitemShut
  {NoStop}%
\bibitem [{\citenamefont {Kornyshev}\ \emph {et~al.}(2014)\citenamefont
  {Kornyshev}, \citenamefont {Luque},\ and\ \citenamefont
  {Schmickler}}]{kornyshev:jsse:2double:14}%
  \BibitemOpen
  \bibfield  {author} {\bibinfo {author} {\bibfnamefont {A.~A.}\ \bibnamefont
  {Kornyshev}}, \bibinfo {author} {\bibfnamefont {N.~B.}\ \bibnamefont
  {Luque}}, \ and\ \bibinfo {author} {\bibfnamefont {W.}~\bibnamefont
  {Schmickler}},\ }\bibfield  {title} {\enquote {\bibinfo {title} {Differential
  capacitance of ionic liquid interface with graphite: the story of two double
  layers},}\ }\href {\doibase 10.1007/s10008-013-2316-8} {\bibfield  {journal}
  {\bibinfo  {journal} {Journal of Solid State Electrochemistry}\ }\textbf
  {\bibinfo {volume} {18}},\ \bibinfo {pages} {1345--1349} (\bibinfo {year}
  {2014})}\BibitemShut {NoStop}%
\bibitem [{\citenamefont {Kaasik}\ \emph {et~al.}(2013)\citenamefont {Kaasik},
  \citenamefont {Tamm}, \citenamefont {Hantel}, \citenamefont {Perre},
  \citenamefont {Aabloo}, \citenamefont {Lust}, \citenamefont {Bazant},\ and\
  \citenamefont {Presser}}]{kaasik2013anisometric}%
  \BibitemOpen
  \bibfield  {author} {\bibinfo {author} {\bibfnamefont {Friedrich}\
  \bibnamefont {Kaasik}}, \bibinfo {author} {\bibfnamefont {Tarmo}\
  \bibnamefont {Tamm}}, \bibinfo {author} {\bibfnamefont {M.~M.}\ \bibnamefont
  {Hantel}}, \bibinfo {author} {\bibfnamefont {E.}~\bibnamefont {Perre}},
  \bibinfo {author} {\bibfnamefont {Alvo}\ \bibnamefont {Aabloo}}, \bibinfo
  {author} {\bibfnamefont {Enn}\ \bibnamefont {Lust}}, \bibinfo {author}
  {\bibfnamefont {M.~Z.}\ \bibnamefont {Bazant}}, \ and\ \bibinfo {author}
  {\bibfnamefont {Volker}\ \bibnamefont {Presser}},\ }\bibfield  {title}
  {\enquote {\bibinfo {title} {Anisometric charge dependent swelling of porous
  carbon in an ionic liquid},}\ }\href@noop {} {\bibfield  {journal} {\bibinfo
  {journal} {Electrochem. Comm.}\ }\textbf {\bibinfo {volume} {34}},\ \bibinfo
  {pages} {196--199} (\bibinfo {year} {2013})}\BibitemShut {NoStop}%
\bibitem [{\citenamefont {Hantel}\ \emph {et~al.}(2014)\citenamefont {Hantel},
  \citenamefont {Weingarth},\ and\ \citenamefont
  {K{\"{o}}tz}}]{hantel2014parameters}%
  \BibitemOpen
  \bibfield  {author} {\bibinfo {author} {\bibfnamefont {M.~M.}\ \bibnamefont
  {Hantel}}, \bibinfo {author} {\bibfnamefont {D.}~\bibnamefont {Weingarth}}, \
  and\ \bibinfo {author} {\bibfnamefont {R.}~\bibnamefont {K{\"{o}}tz}},\
  }\bibfield  {title} {\enquote {\bibinfo {title} {Parameters determining
  dimensional changes of porous carbons during capacitive charging},}\
  }\href@noop {} {\bibfield  {journal} {\bibinfo  {journal} {Carbon}\ }\textbf
  {\bibinfo {volume} {69}},\ \bibinfo {pages} {275--286} (\bibinfo {year}
  {2014})}\BibitemShut {NoStop}%
\bibitem [{\citenamefont {Fic}\ \emph {et~al.}(2010)\citenamefont {Fic},
  \citenamefont {Lot},\ and\ \citenamefont {Frackowiak}}]{fic:ea:10}%
  \BibitemOpen
  \bibfield  {author} {\bibinfo {author} {\bibfnamefont {Krzysztof}\
  \bibnamefont {Fic}}, \bibinfo {author} {\bibfnamefont {Grzegorz}\
  \bibnamefont {Lot}}, \ and\ \bibinfo {author} {\bibfnamefont {Elzbieta}\
  \bibnamefont {Frackowiak}},\ }\bibfield  {title} {\enquote {\bibinfo {title}
  {Electrochemical properties of supercapacitors operating in aqueous
  electrolyte with surfactants},}\ }\href@noop {} {\bibfield  {journal}
  {\bibinfo  {journal} {Electrochim. Acta}\ }\textbf {\bibinfo {volume} {55}},\
  \bibinfo {pages} {7484} (\bibinfo {year} {2010})}\BibitemShut {NoStop}%
\bibitem [{\citenamefont {Fic}\ \emph {et~al.}(2011)\citenamefont {Fic},
  \citenamefont {Lot},\ and\ \citenamefont {Frackowiak}}]{fic:ea:11}%
  \BibitemOpen
  \bibfield  {author} {\bibinfo {author} {\bibfnamefont {Krzysztof}\
  \bibnamefont {Fic}}, \bibinfo {author} {\bibfnamefont {Grzegorz}\
  \bibnamefont {Lot}}, \ and\ \bibinfo {author} {\bibfnamefont {Elzbieta}\
  \bibnamefont {Frackowiak}},\ }\bibfield  {title} {\enquote {\bibinfo {title}
  {Effect of surfactants on capacitance properties of carbon electrodes},}\
  }\href@noop {} {\bibfield  {journal} {\bibinfo  {journal} {Electrochim.
  Acta}\ }\textbf {\bibinfo {volume} {60}},\ \bibinfo {pages} {206} (\bibinfo
  {year} {2011})}\BibitemShut {NoStop}%
\bibitem [{\citenamefont {Lin}\ \emph {et~al.}(2011)\citenamefont {Lin},
  \citenamefont {Taberna}, \citenamefont {Fantini}, \citenamefont {Presser},
  \citenamefont {P{\'{e}}rez}, \citenamefont {Malbosc}, \citenamefont
  {Rupesinghe}, \citenamefont {Teo}, \citenamefont {Gogotsi},\ and\
  \citenamefont {Simon}}]{lin:jpcl:11}%
  \BibitemOpen
  \bibfield  {author} {\bibinfo {author} {\bibfnamefont {Rongying}\
  \bibnamefont {Lin}}, \bibinfo {author} {\bibfnamefont {Pierre-Louis}\
  \bibnamefont {Taberna}}, \bibinfo {author} {\bibfnamefont {S{\'{e}}bastien}\
  \bibnamefont {Fantini}}, \bibinfo {author} {\bibfnamefont {Volker}\
  \bibnamefont {Presser}}, \bibinfo {author} {\bibfnamefont {Carlos~R.}\
  \bibnamefont {P{\'{e}}rez}}, \bibinfo {author} {\bibfnamefont
  {Fran{\c{c}}ois}\ \bibnamefont {Malbosc}}, \bibinfo {author} {\bibfnamefont
  {Nalin~L.}\ \bibnamefont {Rupesinghe}}, \bibinfo {author} {\bibfnamefont
  {Kenneth B.~K.}\ \bibnamefont {Teo}}, \bibinfo {author} {\bibfnamefont
  {Yury}\ \bibnamefont {Gogotsi}}, \ and\ \bibinfo {author} {\bibfnamefont
  {Patrice}\ \bibnamefont {Simon}},\ }\bibfield  {title} {\enquote {\bibinfo
  {title} {Capacitive energy storage from -50 to 100 $^o$c using an ionic
  liquid electrolyte},}\ }\href@noop {} {\bibfield  {journal} {\bibinfo
  {journal} {J. Phys. Chem. Lett.}\ }\textbf {\bibinfo {volume} {2}},\ \bibinfo
  {pages} {2396--2401} (\bibinfo {year} {2011})}\BibitemShut {NoStop}%
\bibitem [{\citenamefont {Hansen}\ and\ \citenamefont
  {MacDonald}(2014)}]{Hansen2014}%
  \BibitemOpen
  \bibfield  {author} {\bibinfo {author} {\bibfnamefont {Jean-Pierre}\
  \bibnamefont {Hansen}}\ and\ \bibinfo {author} {\bibfnamefont {J.~R.}\
  \bibnamefont {MacDonald}},\ }\href@noop {} {\emph {\bibinfo {title} {Theory
  of Simple Liquids}}},\ \bibinfo {edition} {4th}\ ed.\ (\bibinfo  {publisher}
  {Elsevier},\ \bibinfo {year} {2014})\BibitemShut {NoStop}%
\bibitem [{\citenamefont {Kornyshev}\ and\ \citenamefont
  {Schmickler}(1986)}]{kornyshev:jeac:86}%
  \BibitemOpen
  \bibfield  {author} {\bibinfo {author} {\bibfnamefont {Alexej~A.}\
  \bibnamefont {Kornyshev}}\ and\ \bibinfo {author} {\bibfnamefont {Wolfgang}\
  \bibnamefont {Schmickler}},\ }\bibfield  {title} {\enquote {\bibinfo {title}
  {On the coverage dependence of the partial charge transfer coefficient},}\
  }\href@noop {} {\bibfield  {journal} {\bibinfo  {journal} {J. Electroanal.
  Chem.}\ }\textbf {\bibinfo {volume} {202}},\ \bibinfo {pages} {1--21}
  (\bibinfo {year} {1986})}\BibitemShut {NoStop}%
\bibitem [{\citenamefont {Smythe}(1939)}]{smythe:book}%
  \BibitemOpen
  \bibfield  {author} {\bibinfo {author} {\bibfnamefont {William~R.}\
  \bibnamefont {Smythe}},\ }\href@noop {} {\emph {\bibinfo {title} {Static and
  Dynamic Electricity}}}\ (\bibinfo  {publisher} {McGraw Hill},\ \bibinfo
  {year} {1939})\BibitemShut {NoStop}%
\bibitem [{\citenamefont {Kondrat}\ \emph {et~al.}(2013)\citenamefont
  {Kondrat}, \citenamefont {Kornyshev}, \citenamefont {Stoeckli},\ and\
  \citenamefont {Centeno}}]{kondrat:ec:13}%
  \BibitemOpen
  \bibfield  {author} {\bibinfo {author} {\bibfnamefont {S.}~\bibnamefont
  {Kondrat}}, \bibinfo {author} {\bibfnamefont {A.}~\bibnamefont {Kornyshev}},
  \bibinfo {author} {\bibfnamefont {F.}~\bibnamefont {Stoeckli}}, \ and\
  \bibinfo {author} {\bibfnamefont {T.~A.}\ \bibnamefont {Centeno}},\
  }\bibfield  {title} {\enquote {\bibinfo {title} {The effect of dielectric
  constant on capacitance of nanoporous electrodes},}\ }\href@noop {}
  {\bibfield  {journal} {\bibinfo  {journal} {Electrochem. Comm.}\ }\textbf
  {\bibinfo {volume} {34}},\ \bibinfo {pages} {348--350} (\bibinfo {year}
  {2013})}\BibitemShut {NoStop}%
\bibitem [{\citenamefont {Schmidt}\ and\ \citenamefont
  {L{\"{o}}wen}(1996)}]{schmidt1996freezing}%
  \BibitemOpen
  \bibfield  {author} {\bibinfo {author} {\bibfnamefont {Matthias}\
  \bibnamefont {Schmidt}}\ and\ \bibinfo {author} {\bibfnamefont {Hartmut}\
  \bibnamefont {L{\"{o}}wen}},\ }\bibfield  {title} {\enquote {\bibinfo {title}
  {Freezing between two and three dimensions},}\ }\href@noop {} {\bibfield
  {journal} {\bibinfo  {journal} {Phys. Rev. Lett.}\ }\textbf {\bibinfo
  {volume} {76}},\ \bibinfo {pages} {4552} (\bibinfo {year}
  {1996})}\BibitemShut {NoStop}%
\bibitem [{\citenamefont {Schmidt}\ and\ \citenamefont
  {L{\"{o}}wen}(1997)}]{schmidt1997phase}%
  \BibitemOpen
  \bibfield  {author} {\bibinfo {author} {\bibfnamefont {Matthias}\
  \bibnamefont {Schmidt}}\ and\ \bibinfo {author} {\bibfnamefont {Hartmut}\
  \bibnamefont {L{\"{o}}wen}},\ }\bibfield  {title} {\enquote {\bibinfo {title}
  {Phase diagram of hard spheres confined between two parallel plates},}\
  }\href@noop {} {\bibfield  {journal} {\bibinfo  {journal} {Phys. Rev. E}\
  }\textbf {\bibinfo {volume} {55}},\ \bibinfo {pages} {7228} (\bibinfo {year}
  {1997})}\BibitemShut {NoStop}%
\bibitem [{\citenamefont {Holoborodko}(2008)}]{holoborodko:snrd}%
  \BibitemOpen
  \bibfield  {author} {\bibinfo {author} {\bibfnamefont {Pavel}\ \bibnamefont
  {Holoborodko}},\ }\href
  {http://www.holoborodko.com/pavel/numerical-methods/numerical-derivative/smooth-low-noise-differentiators/}
  {\enquote {\bibinfo {title} {Smooth noise robust differentiators},}\
  }\bibinfo {howpublished}
  {http://www.holoborodko.com/pavel/numerical-methods/numerical-derivative/smooth-low-noise-differentiators/}
  (\bibinfo {year} {2008})\BibitemShut {NoStop}%
\end{thebibliography}%

\end{document}